\documentclass[12pt]{article}
\usepackage{a4wide}
\usepackage{epsfig}

\begin{document}
\thispagestyle{empty}
\begin{flushright}
MZ-TH/98-09\\
hep-ph/9806402\\
March 1998
\end{flushright}
\vspace{0.5cm}
\begin{center}
{\Large\bf On the evaluation of sunset-type Feynman diagrams}\\[1truecm]
{\large S.~Groote,$^1$ J.G.~K\"orner$^1$ and A.A.~Pivovarov$^{1,2}$}\\[.7cm]
$^1$ Institut f\"ur Physik, Johannes-Gutenberg-Universit\"at,\\[.2truecm]
  Staudinger Weg 7, D-55099 Mainz, Germany\\[.5truecm]
$^2$ Institute for Nuclear Research of the\\[.2truecm]
  Russian Academy of Sciences, Moscow 117312
\vspace{1truecm}
\end{center}

\begin{abstract}
We introduce an efficient configuration space technique which allows one to 
compute a class of Feynman diagrams which generalize the scalar sunset 
topology to any number of massive internal lines. General tensor vertex 
structures and modifications of the propagators due to particle emission 
with vanishing momenta can be included with only a little change of the 
basic technique described for the scalar case. We discuss applications to 
the computation of $n$-body phase space in $D$-dimensional space-time. 
Substantial simplifications occur for odd space-time dimensions where the 
final results can be expressed in closed form through elementary functions. 
We present explicit analytical formulas for three-dimensional space-time.
\end{abstract}

\newpage

\section{Introduction}
Up to now one has not been able to observe any contradictions to the 
predictions of the Standard Model of particle interactions. Possible 
deviations from the Standard Model or revelations of new physics are 
expected to be quite small at the energies of present accelerators. Future 
experiments focus on tests of the Standard Model with an unprecedented 
precision~\cite{melone1,melone2}. The present accuracy of experimental data 
already demands new levels of accuracy in the theoretical predictions 
of perturbation theory~\cite{melone3}. This requirement leads to the 
necessity to compute multiloop Feynman diagrams beyond the one-loop level 
(as a review, see e.g.~\cite{melone4}). Within the Standard Model the 
diagrams of perturbation theory may contain a multitude of internal and 
external lines with different particles and different masses that makes 
the task of evaluation of such diagrams rather complicated already at the 
two-loop level. The computation of different subsets of diagrams in 
different regimes of their masses and their external momenta is now an 
active field of research and often requires extensive use of direct 
numerical methods~\cite{melone5}.

\begin{figure}\begin{center}
\epsfig{figure=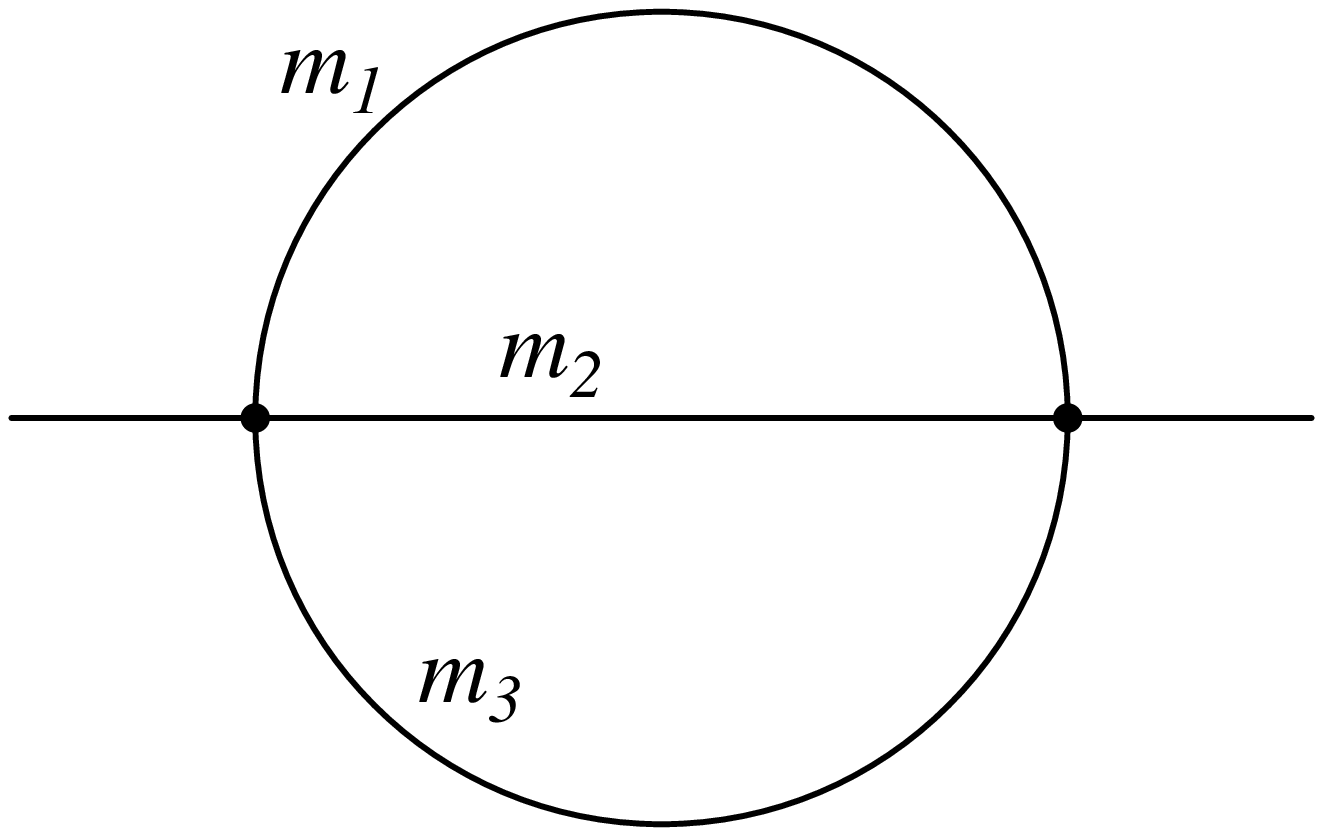, height=3truecm, width=5truecm}\kern2truecm
\epsfig{figure=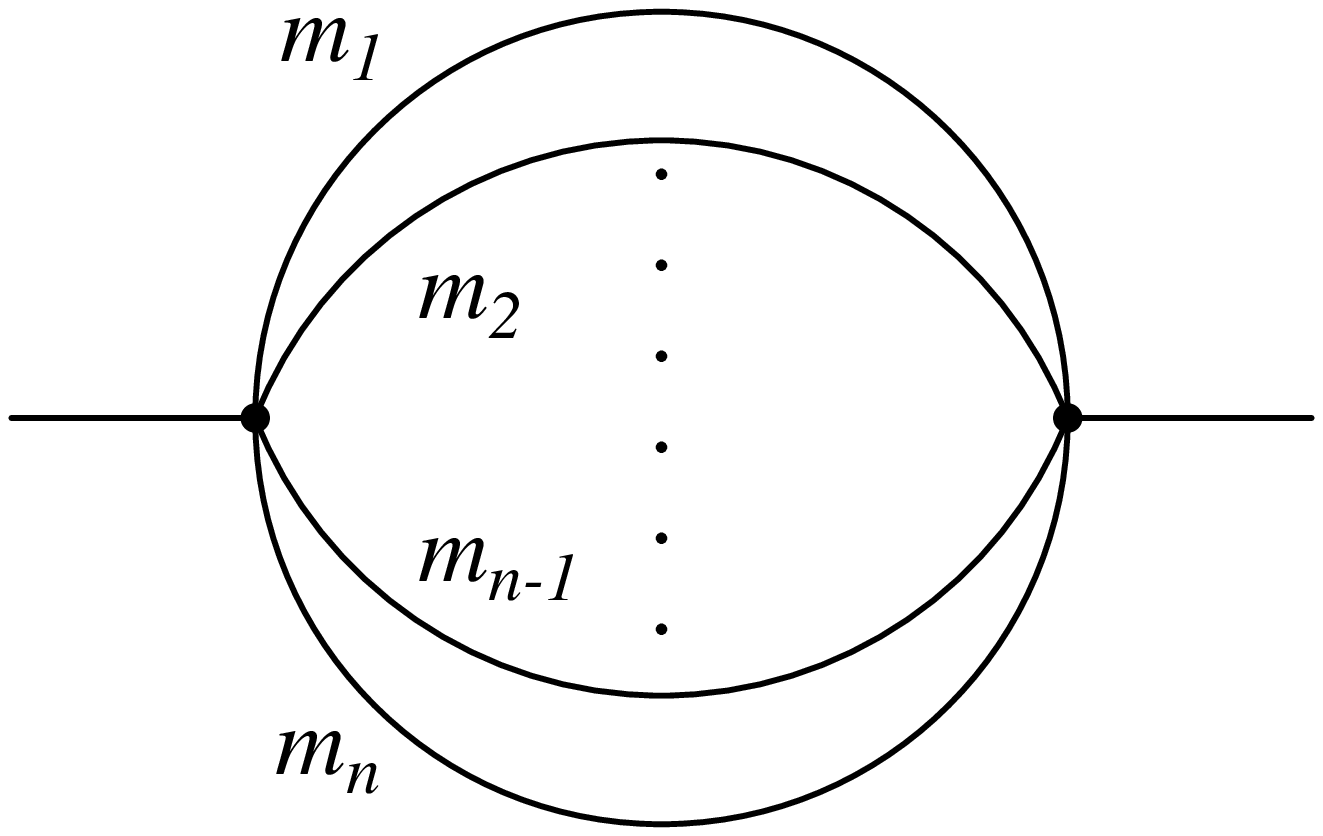, height=3truecm, width=5truecm}
\hbox{\Large\bf\kern4pt(a)\kern6truecm(b)\kern4pt}\break\vspace{12pt}
\epsfig{figure=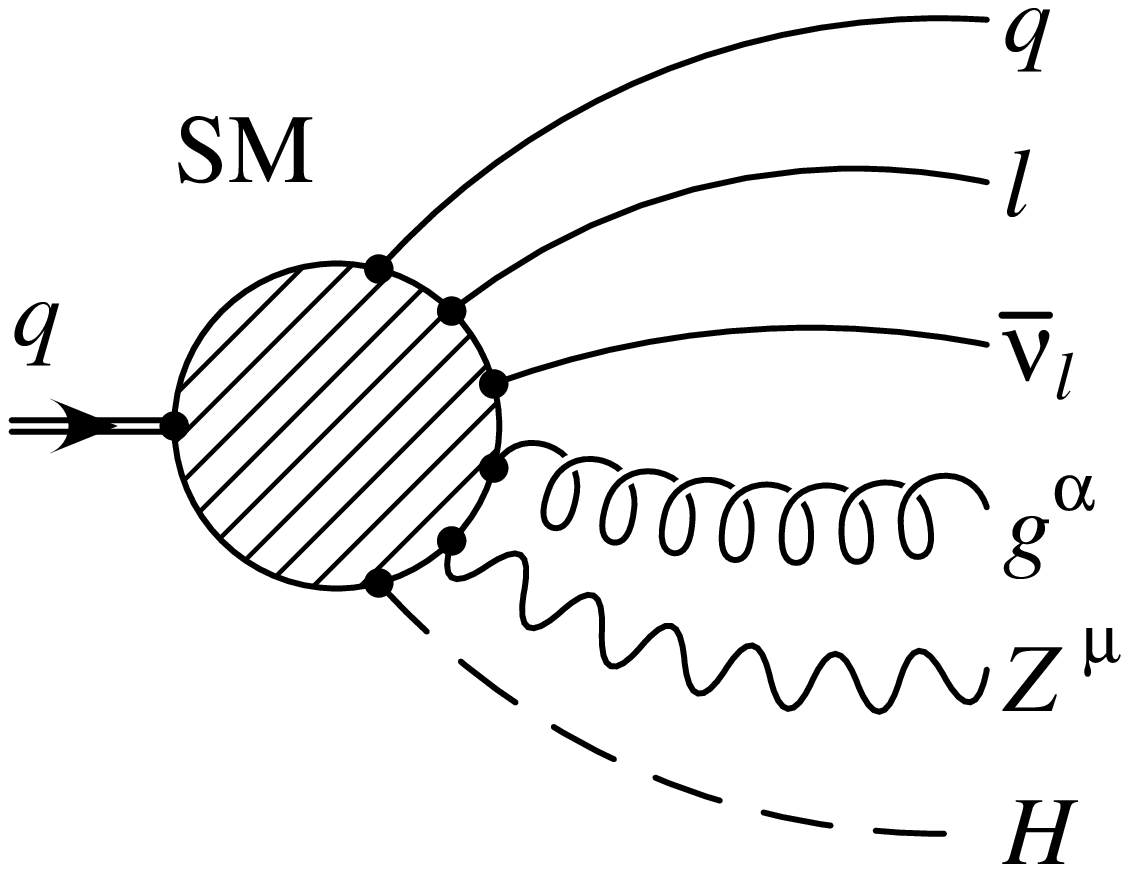, height=5truecm, width=7truecm}\kern12pt
\epsfig{figure=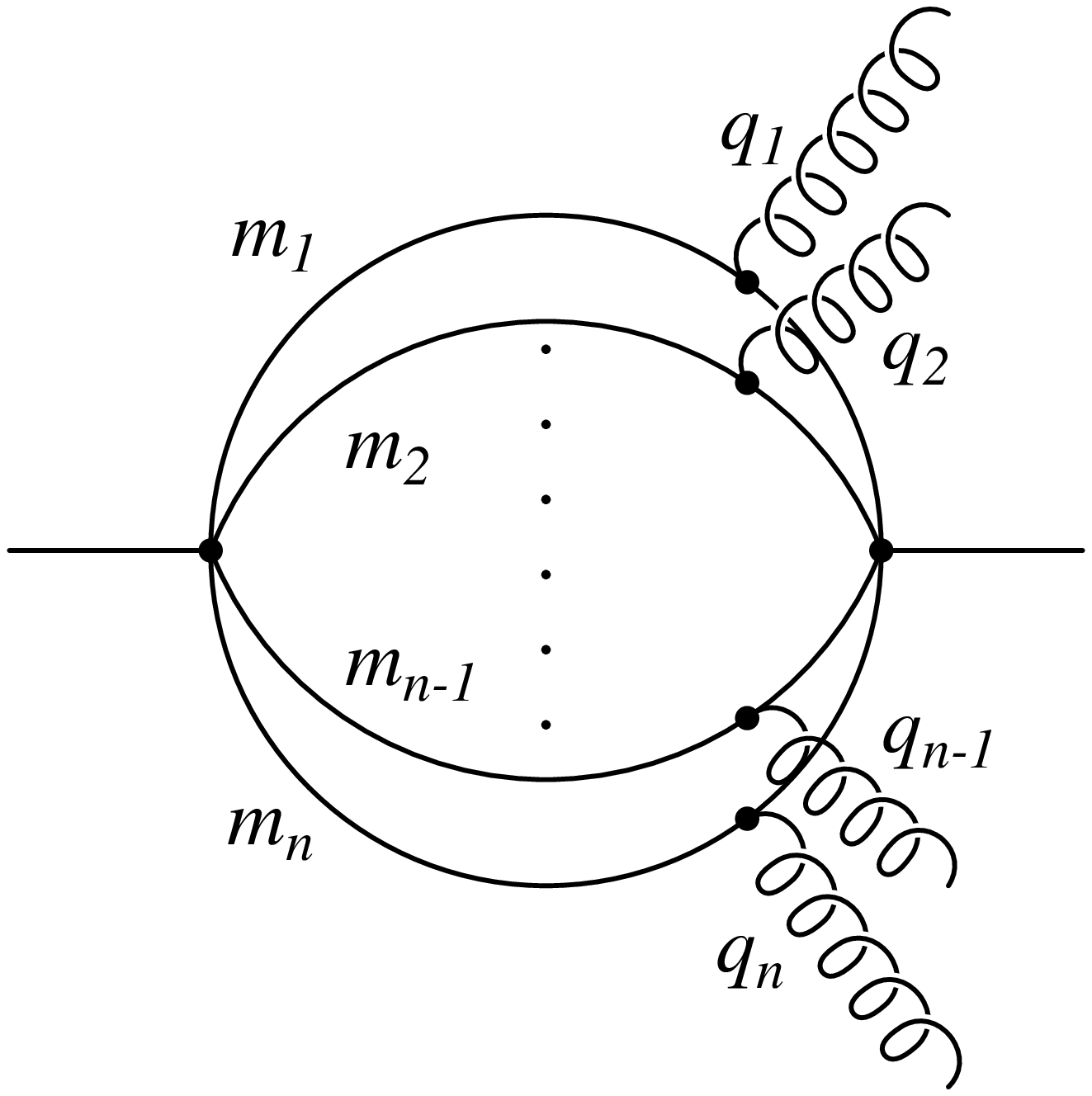, height=5truecm, width=5truecm}
\hbox{\Large\bf\kern4pt(c)\kern6truecm(d)\kern4pt}
\caption{\label{fig1}
(a) a sunset diagram with three different masses $m_1$, $m_2$ and
  $m_3$\hfil\break
(b) general topology of the class of water melon diagrams\hfil\break
(c) one half of a cut $n=6$ water melon diagram in the Standard Model 
representing the spectral function for the process
$q\rightarrow q\ell\bar\nu_\ell gZH$\hfil\break
(d) an example of a water melon diagram with multiple gluon radiation from 
internal lines}
\end{center}\end{figure}

Among the many two-loop topologies the two-loop sunset diagrams with 
different values of internal masses as shown in Fig.~1(a) have been 
recently studied in some detail (see 
e.g.~\cite{melone6,melone7,melone8,melone9,melone10,melone11} and 
references therein). In the present note we describe an efficient method 
for computing and investigating a class of diagrams that generalizes the 
sunset topology to any number of internal lines (massive propagators) in 
arbitrary number of space-time dimensions. We call the class of diagrams 
with this topology water melon diagrams. Fig.~1(b) shows a diagram with 
water melon topology. In our opinion, the method presented in the paper 
completely solves the problem of computing this class of diagrams. The 
method is simple and reduces the multiloop calculation of a water melon 
diagram to a one-dimensional integral which includes only well known 
special functions in the integrand for any values of internal masses. The 
technique is universal and requires only minor technical modifications for 
additional tensor structure of vertices or internal lines (propagators), 
i.e.\ tensor particles or/and fermions can be added at no extra cost. The 
method can also handle form factor type processes at small momentum 
transfer -- the inclusion of lines of incoming/outgoing particles with 
vanishing momenta and derivatives thereof with respect to their momenta is 
straightforward and is done within the same calculational framework. Our 
final one-dimensional integral representation for water melon diagrams is 
well suited for any kind of asymptotic estimates in masses and/or momentum. 
The principal aim of our paper is to work out a practical tool for 
computing water melon diagrams. In the Euclidean domain the numerical 
procedures derived from our representation are efficient and reliable, 
i.e.\ stable against error accumulation. The most interesting part of our 
analysis of water melon diagrams is the construction of the spectral 
decomposition of water melon diagrams, i.e.\ we determine the discontinuity 
across the physical cut in the complex plane of squared momentum. We 
suggest a novel technique for the direct construction of the spectral 
density of water melon diagrams which is based on an integral transform in 
configuration space. We compare our approach to the more traditional way 
of computing the spectral density where one uses an analytic continuation 
in momentum space. Because the analytic structure of water melon diagrams 
is completely fixed by the dispersion representation, our attention is 
focussed on the computation of the spectral density as the basic quantity 
important both for applications and the theoretical investigation of the 
diagram. The complete polarization function can then be easily 
reconstructed from the spectral density with the help of dispersion 
relations. In addition we derive some useful formulas for the polarization 
function in the Euclidean domain and present explicit results for some 
limiting cases where analytical formulas can be found.

\newpage

The paper is organized as follows. In the beginning of  Sec.~2 we describe 
some general properties of water melon diagrams and fix our notation. In 
Sec.~2.1 we introduce the configuration space representation of water melon 
diagrams. In Sec.~2.2 we discuss the ultraviolet (UV) divergence structure 
of water melon diagrams and present the way to regularize the UV 
divergences by subtraction. In Sec.~2.3 some previously known results
are reproduced using our configuration space techniques. Sec.~2.4 contains 
explicit examples in odd-dimensional space-time. In Sec.~2.5 we discuss 
expansions in masses and/or momenta in the Euclidean domain. In Sec.~3 we 
consider the computation of the spectral density of water melon diagrams by
analytic continuation in momentum space. Sec.~4 is devoted to the direct 
computation of the spectral density of water melon diagrams without taking 
recourse to Fourier transforms. Sec.~5 gives our conclusions.

\section{The general framework}
Sunset-type diagrams are two-point functions with $n$ internal propagators 
connecting the initial and final vertex. The sunset diagram shown in 
Fig.~1(a) is the leading order perturbative correction to the lowest 
order propagator in $\phi^4$-theory, i.e.\ it is a two-point two-loop 
diagram with three internal lines. The corresponding leading order 
perturbative correction in $\phi^3$-theory is a one-loop diagram and can be 
considered as a degenerate case of the prior example. A straightforward 
generalization of this topology is a correction to the free propagator in 
$\phi^{n+2}$-theory that contains $n$ loops and $(n+1)$ internal lines (see 
Fig.~1(b)). We call them water melon diagrams\footnote{Alternative names 
that have been suggested in the literature for this topology are banana 
diagrams or basket ball diagrams.}. In a general field theory the internal 
lines may have different masses and may carry different Lorenz structures 
or may contain space-time derivatives. An example of the latter situation 
is a leading quantum correction in higher orders of Chiral Perturbation 
Theory for pseudoscalar mesons where the vertices contain multiple 
derivatives of the meson fields. In order to accommodate such general 
structures we represent a general water melon diagram as a correlator of 
two monomials $j_n(x)$ of the form 
\begin{equation}
j_n(x)={\cal D}_{\mu_1}\phi_1\cdots {\cal D}_{\mu_n}\phi_n  
\end{equation}
where the fields $\phi_n$ have masses $m_n$ and where ${\cal D}_\mu$ is a 
derivative with multi-index $\mu=\{\mu_1,\ldots,\mu_k\}$ standing for 
${\cal D}_\mu=\partial^k/\partial x_{\mu_1}\ldots\partial x_{\mu_k}$. The 
water melon diagrams are contained in the leading order expression for the 
polarization function
\begin{equation}
\Pi(x)=\langle Tj_n(x)j_{n'}(0)\rangle 
\end{equation}
which is explicitly given by a product of propagators and/or their 
derivatives,
\begin{equation}\label{eqn01}
\Pi(x)={\cal D}_{\mu_1\nu_1}(x,m_1)\cdots {\cal D}_{\mu_n\nu_n}(x,m_n).
\end{equation}
Here ${\cal D}_{\mu\nu}(x,m)={\cal D}_\mu {\cal D}_\nu D(x,m)$ is a 
derivative of the propagator $D(x,m)$ with respect to the coordinate $x$ 
with a pair of multi-indices $\{\mu,\nu\}$. The propagator $D(x,m)$ of a 
massive particle with mass $m$ in $D$-dimensional (Euclidean) space-time 
is given by 
\begin{equation}\label{eqn02}
  D(x,m)=\frac1{(2\pi)^D}\int\frac{e^{ip_\mu x^\mu}d^Dp}{p^2+m^2}
  =\frac{(mx)^\lambda K_\lambda(mx)}{(2\pi)^{\lambda+1}x^{2\lambda}}
\end{equation}
where we write $D=2\lambda+2$. $K_\lambda(z)$ is a McDonald function (a 
modified Bessel function of the third kind, see e.g.~\cite{melone12}). The 
propagator $D(x,m)$ depends only on the length of the space-time vector 
$|x|=\sqrt{x_\mu x^\mu}$ for which we simply write $x$. We consider only 
the case $n=n'$ (equal number of lines at the initial and final vertex). We 
thus exclude tadpole configurations (the leaves of this water melon) which 
add nothing interesting. Taking the limit $m\rightarrow 0$ in 
Eq.~(\ref{eqn02}) we get an explicit form of the massless propagator 
\begin{equation}
  D(x,0)=\frac1{(2\pi)^D}\int\frac{e^{ip_\mu x^\mu}d^Dp}{p^2}
  =\frac{\Gamma(\lambda)}{4\pi^{\lambda+1}x^{2\lambda}}
\end{equation}
where $\Gamma(\lambda)$ is the Euler $\Gamma$-function. Note that 
propagators of particles with nonzero spin in configuration space 
representation can be obtained from the scalar propagator by 
differentiation with respect to the space-time point $x$. This does not 
change the functional $x$-structure and causes only minor modification of 
the basic technique. For instance, the propagator for the fermion (spin 1/2 
particle) is given by 
\begin{equation}
S(x,m)=(i\gamma_\mu\frac{\partial}{\partial x_\mu}+m)D(x,m)
\end{equation}
with $\gamma_\mu$ is a Dirac matrix.

The next generalization consists in extending our configuration space 
description to the case of particle radiation from internal lines. The 
modification of the internal line with mass $m$ due to the emission of a 
particle with momentum $q$ at the vertex $V(p,q)$ reads
\begin{eqnarray}\label{eqn03}
D(p,m)&=&{1\over p^2+m^2}\rightarrow {1\over p^2+m^2}V(p,q)
{1\over (p-q)^2+m^2}\nonumber\\  
&\buildrel{q=0}\over=&{1\over p^2+m^2} V(p,0){1\over p^2+m^2}
=-V(p,0){d\over dm^2}{1\over p^2+m^2}.
\end{eqnarray}
The functional $x$-space structure of the corresponding internal line is 
not changed by this modification. Therefore such contributions can be 
obtained either by differentiating the original water melon diagram without 
particle emission with respect to the mass $m$ or by direct differentiation 
of the propagator which leads to a change of the index of the 
corresponding McDonald function. The explicit representation for the 
modified internal line with mass $m$ is given by
\begin{equation}
\label{eqn04}
\frac1{(2\pi)^D}\int\frac{e^{ip_\nu x^\nu}d^Dp}{(p^2+m^2)^{\mu+1}}
  =\frac1{(2\pi)^{\lambda+1}2^\mu\Gamma(\mu+1)}{\left(\frac mx
  \right)^{\lambda-\mu}K_{\lambda-\mu}(mx)}.
\end{equation}
This modified propagator is a product of some power of $x$ and the McDonald 
function as in the standard propagator shown in Eq.~(\ref{eqn02}). The only 
difference is in the index $(\lambda-\mu)$ of the McDonald function which 
is inessential for applications. This does not change the general 
functional structure of the representation constructed below. If there is 
only one-particle emission, the form factor type diagrams with any number 
of massive internal lines can be obtained for any value of $q$ in a closed 
form. One encounters such diagrams when analyzing baryon transitions within 
perturbation theory. The corresponding formulas will be discussed 
elsewhere. Also the change of the mass along the line in Eq.~(\ref{eqn03}) 
is allowed for vanishing incoming/outgoing particle momenta which allows 
for the possibility to discuss processes with radiation of gauge bosons 
such as $b\rightarrow s\gamma$ transition for baryons within the sum rules 
approach. As described above, such generalizations can easily be 
accommodated in our approach without any change in the basic framework. 
From now on we therefore mostly concentrate on the basic scalar master 
configuration of the water melon diagram which contains all features 
necessary for our discussion.

Eq.~(\ref{eqn01}) contains all information about the water melon diagrams 
and in this sense is the final result for the class of diagrams under 
consideration. Of some particular interest is the spectral decomposition 
of the polarization function $\Pi(x)$ which is connected to the particle 
content of a model. As a particular example one can consider a water melon 
diagram with a vertex connecting various particles of the Standard Model 
as shown in Fig.~1(c). The knowledge of the analytic structure of the 
propagators $D(x,m)$ entering Eq.~(\ref{eqn01}) is sufficient for 
determining the analytic structure of the polarization function $\Pi(x)$ 
itself. For applications, however, one may need the Fourier transform of 
the polarization function $\Pi(x)$ given by
\begin{equation}\label{eqn05}
\tilde\Pi(p)=\int\Pi(x)e^{ip_\mu x^\mu}dx
  =\int\langle Tj_n(x)j_{n'}(0)\rangle e^{ip_\mu x^\mu}dx. 
\end{equation}
The tilde ``$\sim$'' on the Fourier transform $\tilde\Pi(p)$ will be 
dropped in the following. The computation of an explicit expression for the 
Fourier transform $\Pi(p)$ for three-line (or two-loop) water melon 
diagrams (the genuine sunset diagram) is described in a number of papers in 
the literature. In the standard, or momentum, representation the quantity 
$\Pi(p)$ is calculated from a $(n-1)$-loop diagram with $(n-1)$ 
integrations over the entangled loop momenta which makes the computation 
difficult when the number of internal lines becomes large. Our technique 
consists in computing the integral in Eq.~(\ref{eqn05}) directly using the 
product of propagators Eq.~(\ref{eqn01}) with their explicit form given by 
Eq.~(\ref{eqn02}). The key simplifying observation is that the rotational 
invariance of the expression in Eq.~(\ref{eqn01}) allows one to perform the 
angular integrations explicitly. As a result one is left with a 
one-dimensional integral over the radial variable in Eq.~(\ref{eqn05}). The 
remaining integrand has a rather simple structure in the form of a product 
of Bessel functions and powers of $x$ which is a convenient starting 
point for further analytical or numerical processing.

\subsection{Configuration space representation}
In the present paper we focus on the technical simplicity and the practical 
applicability of the configuration space approach to the computation of 
water melon diagrams as written down in Eqs.~(\ref{eqn01}), (\ref{eqn02}) 
and~(\ref{eqn05}). The idea of exploiting $x$-space techniques for the 
calculation of Feynman diagrams has a long history. Configuration space 
techniques were successfully used for the evaluation of massless diagrams 
with quite general topologies in~\cite{melone13} and marked a real 
breakthrough in multiloop computations before the invention of the 
integration-by-parts technique.

The case of massive diagrams for general topologies was considered in some 
detail in~\cite{melone14} with $x$-space integration techniques. As it turns 
out, the $x$-space technique has not been very successful for general 
many-loop massive diagrams. The angular integrations do not decouple and no 
decisive simplification occurs. It is the special topology of water melon 
diagrams which makes the $x$-space technique so efficient. Using $x$-space 
techniques one can completely solve the problem of computing this class of 
diagrams.

So, let us taste the water melon. The angular integration in 
Eq.~(\ref{eqn05}) can be explicitly done in $D$-dimensional space-time with 
the result
\begin{equation}\label{eqn06}
\int d^D\hat xe^{ip_\mu x^\mu}=2\pi^{\lambda+1}
  \left(\frac{px}2\right)^{-\lambda}J_\lambda(px)
\end{equation}
where $p=|p|$, $x=|x|$. $J_\lambda(z)$ is the usual Bessel function and 
$d^D\hat x$ is the rotationally invariant measure on the unit sphere in
the $D$-dimensional (Euclidean) space-time. The generalization of 
Eq.~(\ref{eqn06}) to more complicated integrands with additional tensor 
structure $x^{\mu_1}\cdots x^{\mu_k}$ is straightforward and merely leads 
to different orders of the Bessel function after angular averaging. The 
corresponding order of the Bessel function can be easily inferred from the 
expansion of the plane wave function $\exp(ip_\mu x^\mu)$ in a series 
of Gegenbauer polynomials $C^\lambda_j(p_\mu x^\mu/px)$. The 
Gegenbauer polynomials are orthogonal on a $D$-dimensional unit sphere, and 
the expansion of the plane wave $\exp(ip_\mu x^\mu)$ reads
\begin{equation}
\exp(ip_\mu x^\mu)=\Gamma(\lambda)\left(\frac{px}2\right)^{-\lambda}
  \sum_{l=0}^{\infty} i^l (\lambda+l)J_{\lambda+l}(px)
  C^\lambda_l(p_\mu x^\mu/px).
\end{equation}
This formula allows one to single out an irreducible tensorial structure 
from the angular integration in the Fourier integral in Eq.~(\ref{eqn05}). 
Integration techniques involving Gegenbauer polynomials for the computation 
of massless diagrams are described in detail in~\cite{melone13} where many 
useful relations can be found (see also~\cite{melone15}). Our final 
representation of the Fourier transform of a water melon diagram is given 
by the one-dimensional integral
\begin{equation}\label{eqn07}
\Pi(p)=2\pi^{\lambda+1}\int_0^\infty\left(\frac{px}2\right)^{-\lambda}
  J_\lambda(px)D(x,m_1)\cdots D(x,m_n)x^{2\lambda+1}dx
\end{equation}
which is a special kind of integral transformation with a Bessel function 
as a kernel. This integral transformation is known as the Hankel transform. 
The representation given by Eq.~(\ref{eqn07}) is quite universal regardless 
of whether tensor structures are added or particles with vanishing momenta 
are radiated from any of the internal lines. An example of a water melon 
diagram with internal gluon emission is shown in Fig.~1(d).

Next we discuss the analytic structure of a water melon diagram in the 
complex $p$-plane and also the behaviour of $\Pi(p)$ near threshold. From 
Eq.~(\ref{eqn07}) it is clear that $\Pi(p)$ is analytic in the strip 
$|{\rm Im}(p)|<M=\sum_{i=1}^n m_i$. In terms of the relativistically 
invariant variable $p^2$ this means that the function $\Pi(p^2)$ is 
analytic for ${\rm Re}(p^2)>-M^2$ implying that the function $\Pi(p)$ 
becomes singular at the energy $E=M$ in the Minkowskian region. Depending 
on the number of internal lines in the diagram this singularity is either 
a pole for the most degenerate case of only one single propagator or a cut 
in the case of several propagators. 

\subsection{Regularization and subtraction}
The general formula for the Fourier transform (\ref{eqn07}) allows for an 
explicit numerical computation for any momentum $p$ in the Euclidean domain. 
However, if $D>2$ and the number of propagators is sufficiently large, the 
integral diverges in the ultraviolet (UV) region or, equivalently, at small 
$x$. Note that for $D=2$ there are only logarithmic singularities at the 
origin ($\lambda=0$ for the propagator Eq.~(\ref{eqn02})) which  makes this 
case technically simpler. Also for $D=2$ the strength of the singularity 
does not increase with the number of internal lines as dramatically as for 
higher dimensions. In the general case the UV divergence prevents one from 
taking the limit $D\rightarrow D_0$ where $D_0$ is an integer number of 
physical space-time dimensions. Taken by itself, the representation given 
by Eq.~(\ref{eqn07}) determines the dimensionally regularized function for 
complex space-time dimension $D$. For the scalar propagator in space-time 
with $D_0=4$ the singularity at small $x$ is given by $x^{-2}$. For 
explicit computations in this paper we normally use dimensional 
regularization, although our particular way of regularization is not always 
the orthodox one. The structure of the UV divergence is very simple for the 
general water melon diagram. Any water melon diagram only has an overall 
divergence without subdivergences if the $R$-operation using normal 
ordering and vanishing tadpoles (see e.g.~\cite{melone16}) is properly 
defined which means that the water melon is stripped off its leaves. Thus 
the renormalization of water melon diagrams in the representation given 
by Eq.~(\ref{eqn07}) is simple and well suited for a numerical treatment 
which is important for practical applications.

Besides dimensional regularization the momentum space subtraction technique 
is sometimes used for dealing with UV divergences. Its implementation is 
very simple in the representation given by Eq.~(\ref{eqn07}). Subtractions 
at the origin $p=0$ are always possible as long as one has massive internal 
lines which prevent the appearance of infrared singularities. If this is 
the case, the subtraction amounts to expanding the function 
\begin{equation}
\label{eqn08}
\left(\frac{px}2\right)^{-\lambda}J_\lambda(px),
\end{equation}
(which is a kernel or weight function of the integral transformation in 
Eq.~(\ref{eqn07})) in a Taylor series around $p=0$ in terms of a polynomial 
series in $p^2$. The order $N$ subtraction is achieved by writing 
\begin{equation}\label{eqn09}
\left[\left(\frac{px}2\right)^{-\lambda}J_\lambda(px)\right]_N
  =\left(\frac{px}2\right)^{-\lambda}J_\lambda(px)
  -\sum_{k=0}^N\frac{(-1)^k}{k!\Gamma(\lambda+k+1)}
  \left(\frac{px}2\right)^{2k}
\end{equation}
and by keeping $N$ terms in the expansion on the right hand side. 
Substituting expansion Eq.~(\ref{eqn09}) into expression (\ref{eqn07})
leads to a momentum subtracted polarization function
\begin{equation}\label{eqn10}
\Pi_{\rm Mom}(p)=\Pi(p)-\sum_{k=0}^N
  \frac{p^{2k}}{k!}\left(\frac{d}{d p^2}\right)^k \Pi(p)|_{p^2=0}
\end{equation}
which is finite if the number of subtractions is sufficiently high. The 
function $\Pi(p)$ itself is divergent as well as any derivative on the
right hand side of Eq.~(\ref{eqn10}) and requires regularization. However, 
the difference or the quantity $\Pi_{\rm Mom}(p)$ is finite and independent 
of any regularization used to give a meaning to the individual terms in 
Eq.~(\ref{eqn10}). Note that the expansion (\ref{eqn09}) is a polynomial 
in $p^2$ in accordance with the general structure of the 
$R$-operation~\cite{melone16}. The number $N$ of necessary subtractions is 
determined by the divergence index of the diagram and can be found 
according to the standard rules~\cite{melone16}. The subtraction at the 
origin $p=0$ is allowed if there is at least one massive line in the 
diagram and an arbitrary number of massless lines. If there are no massive 
internal lines, the corresponding diagram can easily be calculated 
analytically and the problem of subtraction is trivial. After having 
performed the requisite subtraction one can take the limit 
$D\rightarrow D_0$ in Eq.~(\ref{eqn07}) where $D_0$ is an integer. The 
diagram as a whole becomes finite after the subtraction. In order to obtain 
a deeper insight and for reasons of technical convenience it is useful to 
give a meaning to every individual term in the expansion in 
Eq.~(\ref{eqn07}) after substituting the difference given by 
Eq.~(\ref{eqn09}) which will then finally lead to Eq.~(\ref{eqn10}). 
To make the individual terms meaningful one has to introduce an 
intermediate regularization. This intermediate regularization can in 
principle be the standard dimensional regularization. However, the 
regularization in this particular case can be also achieved by adding a 
factor $x^{2\omega}$ to the integration measure which, in practice, turns 
into $(\mu x)^{2\omega}$ in order to keep the overall dimensionality of 
the diagram correct, where $\mu$ is an arbitrary mass parameter. Thus, all 
propagators are taken in the form of an integer number of dimensions 
(depending on the choice of the space-time) but the integration measure is 
modified to provide regularization. We refer to this procedure of auxiliary 
regularization as an unorthodox dimensional regularization~\cite{melone17}.
Note that similar modifications of dimensional regularization are known in 
other applications. For instance, in some supersymmetric theories one has
to keep the four dimensional structure of tensor fields to preserve the 
Ward identities for the regularized theory. The corresponding modification 
of the standard dimensional regularization is then called dimensional 
regularization by dimension reduction.

Let us now demonstrate that finite quantities (like the momentum 
subtracted polarization operator in Eq.~(\ref{eqn10})) are independent of 
the intermediate regularization scheme used. Consider the simplest case of 
one massive and one massless line. The corresponding diagram requires only 
one subtraction according to the standard power counting by noting that its 
divergence index is equal to zero. 
Within ordinary dimensional regularization the polarization function reads 
\begin{eqnarray}
\Pi^D(p)&=&\int D(x,m) D(x,0)e^{ip_\mu x^\mu}d^Dx\nonumber\\
  &=&{\Gamma(1-\lambda)m^{2\lambda -2} \over (4\pi)^{\lambda+1}\lambda}
  \,{}_2F_1\left(1,1-\lambda;\lambda+1;-\frac{p^2}{m^2}\right)
\end{eqnarray}
where ${}_2F_1(a,b;c;z)$ is a hypergeometric function and 
$\lambda=1-\varepsilon$. Within the unorthodox dimensional regularization
we find
\begin{eqnarray}
\Pi_{4,\omega}(p)&=&\int D_4(x,m)D_4(x,0)e^{ip_\mu x^\mu}x^{2\omega}d^4x
  \nonumber\\
  &=&\frac{\Gamma(1+\omega)\Gamma(\omega)(m^2/4)^{-\omega}}{(4\pi)^2}
  \,{}_2F_1\left(1+\omega,\omega;2;-\frac{p^2}{m^2}\right).
\end{eqnarray}
It is straightforward to see that in the limits $\omega\rightarrow 0$ and 
$\varepsilon\rightarrow 0$ both $\Pi^D(p)-\Pi^D(0)$ and 
$\Pi_{4,\omega}(p)-\Pi_{4,\omega}(0)$ are finite and equal to each other
in accordance with Eq.~(\ref{eqn10}) and general statements of the 
$R$-operation.

The unorthodox dimensional regularization scheme is only an intermediate 
step in the calculation of finite quantities. However, in some cases it may 
lead to dramatic simplifications in the calculation of the relevant 
integrals. As we shall see later on, the computation of the limit 
$D\rightarrow D_0=3$ (and of any odd-dimensional space-time) can be 
explicitly done for any water melon diagram within unorthodox dimensional 
regularization because of the simple form of the propagators and the 
weight function: they contain the exponential 
functions instead of Bessel functions. 
The use of ordinary dimensional regularization would require the 
full computation for the noninteger space-time dimension $D$ first which is 
not possible for an arbitrary diagram.

\subsection{Test of the technique with known results}
In this subsection we reproduce some known results with our configuration 
space techniques. The complexity of the examples increases with the number 
of massive propagators occuring in the water melon diagram. The completely 
massless case is quite trivial and will not be discussed futher. Water 
melon diagrams with one massive line and an arbitrary number of massless 
lines are solved with the formula 
\begin{equation}
\int_0^\infty x^\mu K_\nu(mx)dx
=2^{\mu-1}m^{-\mu-1}\Gamma\left(\frac{1+\mu+\nu}2\right)
\Gamma\left(\frac{1+\mu-\nu}2\right)
\end{equation}
which allows one to compute all necessary counterterms. Water melon 
diagrams containing two massive lines and arbitrary number of massless 
lines are also solved in a closed analytical form with the formula 
\begin{equation}
\label{eqn11}
\int_0^\infty x^{2\alpha-1} K_\mu(mx) K_\mu(m x)dx
={2^{2\alpha-3}\over m^{2\alpha}\Gamma(2\alpha)}
\Gamma(\alpha+\mu)\Gamma(\alpha)
\Gamma(\alpha)\Gamma(\alpha-\mu).
\end{equation}
There exists a generalization of Eq.~(\ref{eqn11}) to different masses 
and indices of the McDonald functions as well. We do not dwell upon this 
point here. As a simple example we give the result for a three-loop water 
melon diagram with two massive and two massless lines at vanishing external 
momentum. The analytical expression for the diagram in the configuration 
space representation is
\begin{equation}\label{eqn12}
  \Pi(0)=\int D(x,m)^2D(x,0)^2d^Dx=\int
  \left(\frac{(mx)^\lambda K_\lambda(mx)}{(2\pi)^{\lambda+1}x^{2\lambda}}
  \right)^2\left(\frac{\Gamma(\lambda)}{4\pi^{\lambda+1}x^{2\lambda}}
  \right)^2 d^Dx.
\end{equation}
While the angular integration in $D$-dimensional space-time is trivial 
the problem of residual radial integration is solved by Eq.~(\ref{eqn11}).
The result for the integral in Eq.~(\ref{eqn12}) is
\begin{equation}
\Pi(0)=\left(\frac{m^2}{4}\right)^{3\lambda-1}\frac1{2^8\pi^{3\lambda+3}}
\frac{\Gamma(\lambda)^2\Gamma(1-\lambda)\Gamma(1-2\lambda)^2
  \Gamma(1-3\lambda)}{\Gamma(\lambda+1)\Gamma(2-4\lambda)}.
\end{equation}
This result corresponds to the quantity $M_1$ which is the simplest basis 
element for the computation of massive three-loop diagrams in a general 
three-loop topology considered in~\cite{melone18}.

Next we turn to a sunset diagram with three massive lines, i.e.\ the 
two-loop water melon diagram. There is an analytical expression for such 
a diagram at some special values of external momenta computed within 
dimensional regularization~\cite{melone10}. We reproduce this result here. 
We start with Eq.~(\ref{eqn05}) setting $n=n'=3$. The angular integration 
just gives the volume of $D$-dimensional sphere. We then arrive at 
Eq.~(\ref{eqn07}) with $n=3$. Therefore the one-dimensional integral to 
analyze is 
\begin{equation}
\Pi_D(p)=2\pi^{\lambda+1}\int_0^\infty\left(\frac{px}2\right)^{-\lambda}
  J_\lambda(px)D(x,m_1)D(x,m_2) D(x,m_3)x^{2\lambda+1}dx.
\end{equation}
To localize the finite part we first use momentum subtraction and
separate $\Pi_D(p)$ into its finite and infinite (but dimensionally
regularized) parts
\begin{equation}
\Pi_D(p)=\Pi_{\rm Mom}(p)+\Pi_{\rm sing}(p)
\end{equation}
where $\Pi_{\rm Mom}(p)$ is a momentum subtracted polarization function 
and $\Pi_{\rm sing}(p)$ is a counterterm in dimensional regularization.
Only two subtractions are necessary by power counting. The explicit 
expression for the momentum subtracted polarization function is
\begin{eqnarray}
\Pi_{\rm Mom}(p)&=&2\pi^{\lambda+1}\int_0^\infty
  \left[\left(\frac{px}2\right)^{-\lambda}J_\lambda(px)\right]_1
  D(x,m_1)D(x,m_2) D(x,m_3)x^{2\lambda+1}dx\nonumber\\
  &=&2\pi^{\lambda+1}\int_0^\infty\left[\left(\frac{px}2\right)^{-\lambda}
  J_\lambda(px)-\frac1{\Gamma(\lambda+1)}
  +\frac{p^2x^2}4\frac1{\Gamma(\lambda+2)}\right]\nonumber\\&&\qquad
  \times D(x,m_1)D(x,m_2) D(x,m_3)x^{2\lambda+1}dx.
\end{eqnarray}
The singular part is given by a first order polynomial in $p^2$ with 
coefficients $A$, $B$ whose values depend on the regularization scheme (in 
this case dimensional regularization is used)
\begin{eqnarray}\label{eqn13}
\Pi_{\rm sing}(p)&=&A+p^2B\nonumber\\
  &=&\frac{2\pi^{\lambda+1}}{\Gamma(\lambda+1)}\int_0^\infty D(x,m_1)D(x,m_2)
  D(x,m_3)x^{2\lambda+1}dx\nonumber\\&&
  -p^2\frac{2\pi^{\lambda+1}}{4\Gamma(\lambda+2)}\int_0^\infty 
  x^2D(x,m_1)D(x,m_2) D(x,m_3)x^{2\lambda+1}dx.
\end{eqnarray}
With this representation our strategy applies straightforwardly. In the 
momentum subtracted part one can forego the regularization (it is finite by 
$R$-operation) and perform the one-dimensional integration numerically for 
$D=4$. The counterterms are then simple numbers independent of $p$. They 
contain divergent parts (regularized within dimensional regularization) and 
need to be computed only once to recover the function $\Pi(p)$ for any $p$.

In the particular case of the sunset diagram (three lines) the necessary 
integrals are known analytically in their full form and can be found in 
integral tables (e.g.~\cite{melone19}). Using the tables may not always be 
convenient and we therefore present a simplified approach which allows one 
to deal even with the complicated cases in a simpler manner. Let us specify 
to the particular case $p=m_1+m_2-m_3$ (pseudothreshold) where an analytical 
answer exists~\cite{melone10}. For simplicity we choose $m_1=m_2=m_3/2=m$. 
Then $p=0$ and $\Pi_{\rm Mom}(0)=0$ (this is a regular function at this 
Euclidean point). In this case the counterterm $p^2B$ vanishes because at 
finite $\varepsilon$ the quantity $B$ is finite. We thus only have to 
consider $A$. On the other hand our considerations are completely general 
since one only requires more terms in the $p^2$-expansion for arbitrary 
$p$. For the special mass configuration considered here the result 
of~\cite{melone10} reads
\begin{equation}\label{eqn14}
\Pi_D^{ref}(0)=\pi^{4-2\varepsilon}\frac{m^{2-4\varepsilon}
  \Gamma^2(1+\varepsilon)}{(1-\varepsilon)(1-2\varepsilon)}
  \left[-\frac3{\varepsilon^2}+\frac{8\ln 2}{\varepsilon}
  -8\ln^2 2\right]+O(\varepsilon).
\end{equation}
This result can be extracted from the first line of Eq.~(\ref{eqn13})
using the integral tables given in~\cite{melone19}. However, in the general 
mass caes the necessary formulas are rather cumbersome. Even for the 
special mass configuration considered here they are not so simple. We 
therefore discuss a short cut which allows one to obtain results 
immediately without having to resort to integral tables. What we really 
only need is an expansion in $\varepsilon$. Basically we need the integral 
\begin{equation}
\int_0^\infty D(x,m)D(x,m)D(x,2m)x^{2\lambda+1}dx
\end{equation} 
which is of the general form 
\begin{equation}\label{eqn15}
\int_0^\infty x^\rho K_\mu(mx)K_\mu(m x)K_\mu(2mx)dx  
\end{equation}
($\mu=\lambda$ and $\rho=1-\lambda$ in our case). For two McDonald functions 
in the integrand (without the last one in the above equation, for instance)
the result is given by Eq.~(\ref{eqn11}). Let us reduce the problem at hand 
to Eq.~(\ref{eqn11}) and do our numerical evaluations with functions in 
four-dimensional space-time where no regularization is necessary. To do 
this we subtract the leading singularities at small $\xi$ from the last 
McDonald function in Eq.~(\ref{eqn15}) using the series expansion near 
the origin,
\begin{equation}
\left(\frac\xi2\right)^\lambda K_\lambda(\xi)=
\frac{\Gamma(\lambda)}2\left[1+\frac1{1-\lambda}\left(\frac\xi2\right)^2
-\frac{\Gamma(1-\lambda)}{\Gamma(1+\lambda)}
  \left(\frac\xi2\right)^{2\lambda}\right]+O(\xi^4,\xi^{2+2\lambda}).
\end{equation}
After decomposing the whole answer into finite and singular parts as 
$(2\pi)^{2D}A=F+S$ (where the total normalization of~\cite{melone10} has 
been adopted in the definition of $F$ and $S$) we find for the singular 
part
\begin{eqnarray}\label{eqn16}
\lefteqn{S\ =\ \frac{(2\pi)^D m^{2\lambda}}{\Gamma(\lambda+1)}
  \int_0^\infty x^{2(1-\lambda)-1} K_\lambda(mx)K_\lambda(m x)}\nonumber\\&&
  \times\frac{\Gamma(\lambda)}2\left[1+\frac{(mx)^2}{1-\lambda}
  -\frac{\Gamma(1-\lambda)}{\Gamma(1+\lambda)}(mx)^{2\lambda}\right]dx
  \nonumber\\
&=&\frac{(2\pi)^D m^{2-4\varepsilon}}{\Gamma(\lambda+1)}\int_0^\infty
  \xi^{2\varepsilon-1}K_\lambda(\xi)K_\lambda(\xi)\frac{\Gamma(\lambda)}2
  \left[1+\frac{\xi^2}{1-\lambda}-\frac{\Gamma(1-\lambda)}{\Gamma(1+\lambda)}
  \xi^{2\lambda}\right]d\xi\nonumber\\
&=&\pi^{4-2\varepsilon}\frac{m^{2-4\varepsilon}\Gamma^2(1+\varepsilon)}
  {(1-\varepsilon)(1-2 \varepsilon)}\left[-\frac3{\varepsilon^2}
  +\frac{8\ln 2}{\varepsilon}+8(2-2\ln 2-\ln^2 2)\right]+O(\varepsilon).
\end{eqnarray}
The pole contributions coincide with the result in Eq.~(\ref{eqn14}) while 
the finite part is different. It is corrected by the finite expression
\begin{eqnarray}
F&=&\frac{(2\pi)^Dm^{2\lambda}}{\Gamma(\lambda+1)}
  \int_0^\infty x^{2(1-\lambda)-1}K_\lambda(mx)K_\lambda(m x)\\&&
  \times\left\{(mx)^\lambda K_\lambda(2mx)
  -\frac{\Gamma(\lambda)}2\left[1+\frac{(mx)^2}{1-\lambda}
  -\frac{\Gamma(1-\lambda)}{\Gamma(1+\lambda)}(mx)^{2\lambda}\right]
  \right\}dx.\nonumber
\end{eqnarray}
Because this quantity is finite (no strong singularity at small $x$) one 
can put $D=4$ to get
\begin{equation}
F=16\pi^4m^2\int_0^\infty\frac{dx}xK_1(x)K_1(x)
  \left\{xK_1(2x)-\frac12\left[1+x^2(-1+2\gamma_E+2\ln x)\right]\right\}
\end{equation}
where $\gamma_E$ is the Euler's constant gamma with the numerical value 
$0.577\ldots$. The numerical integration results in
\begin{equation}
F=16\pi^4m^2[-0.306853\ldots\ ]=16\pi^4m^2[-(1-\ln 2)\times 1.0000\ldots\ ].
\end{equation}
Now one can restore all $\varepsilon$ dependence in the normalization 
factors as in Eqs.~(\ref{eqn14}) and~(\ref{eqn16}) because $F$ is not 
singular in $\varepsilon$ and this change of normalization is absorbed in 
the $O(\varepsilon)$ symbol. One obtains
\begin{equation}\label{eqn17}
F=\pi^{4-2\varepsilon}\frac{m^{2-4\varepsilon}\Gamma^2(1+\varepsilon)}
  {(1-\varepsilon)(1-2 \varepsilon)}
  \left[-16(1-\ln 2)\times 1.0000\ldots\ \right]+O(\varepsilon).
\end{equation}

Adding $F$ from Eq.~(\ref{eqn17}) to $S$ from Eq.~(\ref{eqn16}) one 
obtains the result in Eq.~(\ref{eqn14}). Of course we use the
analytical expression $(1-\ln 2)$ 
for illustrative reasons because we know the final 
answer.

We emphasize that there is nothing new in computing the polarization 
function related to this diagram at any $p$. Just some finite part appears 
(from the momentum subtracted polarization function). Also one needs 
another counterterm for non-zero $p^2$. Its computation is analogous to what 
has been done here. It is even simpler because the singularity at small $x$ 
is weaker and only one subtraction from the McDonald function is necessary.
This technique works for computation at any complex $p^2$.

We add a general remark about the computation near a threshold. The value 
of the polarization function at threshold was obtained in~\cite{melone10}. 
However, the polarization function is not analytic in the squared external 
momentum near threshold so it cannot be expanded in a regular Taylor series. 
Some high order derivatives of the polarization function considered as a 
function of the real variable $p^2$ do not exist at the threshold. While we 
do get numerical improvement until the finite derivatives are available 
(because the very singularity is suppressed in these orders), the analytic 
structure of the singularity cannot be obtained in the way of a series 
expansion.

The examples presented in this subsection so far are well known and have 
been obtained before using techniques differing from ours. While we can 
numerically compute any water melon diagram with any arbitrary number of 
internal massive lines it is not easy to find corresponding analytical 
expressions for comparison in the literature. Beyond two loops there are 
only few examples in the literature. We have discovered a relevant result 
in~\cite{melone18} where an efficient technique for the computation of 
three-loop massive diagrams of a general topology has been 
developed\footnote{We thank David Broadhurst for an illuminative discussion 
of this point}. Our test case corresponds to the quantity $B_N(0,0,2,2,2,1)$ 
in~\cite{melone18} and is a three-loop massive water melom diagram at
vanishing external momentum (see Fig.~2).
\begin{figure}\begin{center}
\epsfig{figure=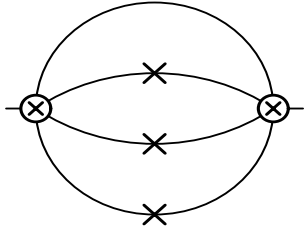, height=3truecm, width=4truecm}
\caption{\label{fig2}Three loop massive water melon diagram with propagator
doubling. The cross on the lines corresponds to one differentiation with
respect to the propagator mass}
\end{center}\end{figure}
Three of the four propagators of the diagram are differentiated in the mass,
i.e.\ the analytical expression for the propagator in momentum space is
$(p^2+m^2)^{-2}$. The reference result of~\cite{melone18} reads
\begin{equation}\label{eqn18}
B_N(0,0,2,2,2,1)=-\frac{7}{4}\zeta(3).
\end{equation}
The representation in configuration space for such a diagram (after 
rotation to Euclidean space-time) has the form
\begin{equation}\label{eqn19}
{\rm wm}4=-4\int_0^\infty K_0(x)^3K_1(x)x^2dx
\end{equation}
where the explicit expression for the derivative of the propagator in 
Eq.~(\ref{eqn04}) with respect to the mass has been used. The diagram is 
finite (that was the reason for differentiation of the propagators). 
Therefore the expression for space-time dimension $D=4$ is used. Numerical 
integration of Eq.~(\ref{eqn19}) gives
\begin{equation}
{\rm wm}4=-2.1036\ldots=-\frac{7}{4}\zeta(3)\times[1.0000\ldots\ ]
\end{equation}
which coincides with Eq.~(\ref{eqn18}).

We conclude that the configuration space method is simple and rather 
efficient for numerical computation of diagrams with water melon topology. 
When the structure of the transcendentality of the result is known for a 
particular diagram from some other considerations (as in the latest example 
where only $\zeta(3)$ is present) our numerical technique can be used in 
some cases to restore the rational coefficients of these 
transcendentalities which can be thought of as elements of the basis for a 
class of diagrams. 
 
\subsection{Examples in odd-dimensional space-time}
It is interesting to note that the computation of Eq.~(\ref{eqn07}) can be 
performed in closed form for any number of internal lines in space-times 
of odd dimensions. As the simplest example we take three-dimensional 
space-time $D\rightarrow D_0=3$.

For $\lambda_0=(D_0-2)/2=1/2$ with $D_0=3$ the propagator in 
Eq.~(\ref{eqn02}) reads
\begin{equation}
D(x,m)\rightarrow
  D_3(x,m)=\frac{\sqrt{mx}K_{1/2}(mx)}{(2\pi)^{3/2}x}=\frac{e^{-mx}}{4\pi x}
\end{equation}
while the weight function after the angular integration given by 
Eq.~(\ref{eqn08}) at $\lambda=\lambda_0=1/2$ becomes
\begin{equation}
\left(\frac{px}2\right)^{-1/2}J_{1/2}(px)
  =\frac2{\sqrt{\pi}}\frac{\sin(px)}{px}.
\end{equation}
The explicit result for the $n$-line water melon diagram is then given by the 
integral
\begin{eqnarray}\label{eqn20}
\Pi(p)&=&4\pi\int_0^\infty\frac{\sin(px)}{px}\frac{e^{-Mx}}{(4\pi x)^{n-2}}
  (\mu x)^{2\epsilon}dx\nonumber\\
&=&{\Gamma(2-n+2\epsilon)\over 2ip(4\pi)^{n-1}}
\left[(M-ip)^{n-2-2\epsilon}-(M+ip)^{n-2-2\epsilon}\right]\mu^{2\epsilon}
\end{eqnarray}
where $\epsilon$ is used for regularization and $M=\sum m_i$.

Here we see the advantage of using 
the unorthodox dimensional regularization
for computing the finite part of water melon diagrams.
The essential simplification 
of the functional form of the integrand 
within unorthodox dimensional regularization
in odd-dimensional space-time 
allows one to compute any water melon diagram
analytically.

We consider some particular cases of Eq.~(\ref{eqn20}) for different 
values of $n$. For $n=1$ we simply recover the propagator function with 
the discontinuity 
\begin{equation}
\rho(s)=\frac{\mbox{Disc\,}\Pi(p)}{2\pi i}=\frac1{2\pi i}\left(\Pi(p)
  \Big|_{p^2=s\exp(-i\pi)}-\Pi(p)\Big|_{p^2=s\exp(i\pi)}\right)=\delta(s-m^2)
\end{equation}
where $s$ is the squared energy, $s=-p^2$. It is usual to call this 
expression the spectral density associated with the diagram. For $n=2$ the 
answer for the polarization function $\Pi(p)$ is still finite 
(no regularization is required) and is given by
\begin{equation}\label{eqn21}
\Pi(p)=\frac1{8\pi ip}\ln\left(\frac{M+ip}{M-ip}\right).
\end{equation}
The spectral density, i.e.\ the discontinuity of Eq.~(\ref{eqn21}) across 
the cut in the energy square complex $p^2$ plane is given by  
\begin{equation}\label{eqn22}
\rho(s)={1\over 8\pi\sqrt{s}}\theta(s-(m_1+m_2)^2),\qquad s=-p^2,\quad s>0 
\end{equation}
which is nothing but three-dimensional two-particle phase space. This can 
be immediately checked by direct computation. The cases with $n>2$ have 
more structure and therefore are more interesting. For the proper sunset 
diagram with $n=3$, Eq.~(\ref{eqn20}) leads to 
\begin{equation}\label{eqn23}
\Pi(p)=\frac1{32\pi^2}\left(\frac1\epsilon-\frac{M}{ip}\ln\left(
  \frac{M+ip}{M-ip}\right)-\ln\left(\frac{M^2+p^2}{\mu^2}\right)\right).  
\end{equation}
The arbitrary scale $\mu^2$ appears due to our way of regularization. 
However, the discontinuity of the polarization function in Eq.~(\ref{eqn23}) 
is independent of $\mu^2$ as it must be.
The discontinuity is finite and, therefore, independent of the 
regularization used. Accordingly, Eq.~(\ref{eqn23}) has the correct 
spectral density,
\begin{equation}
\rho(s)=\frac{\sqrt s-M}{32\pi^2\sqrt s}\,\theta(s-M^2).
\end{equation}
The general formula for the spectral density for any $n>1$ in $D=3$ can be 
extracted from Eq.~(\ref{eqn20}). It reads
\begin{equation}\label{eqn24}
\rho(s)=\frac{(\sqrt s-M)^{n-2}}{2(4\pi)^{n-1}(n-2)!\sqrt s}\,\theta(s-M^2).  
\end{equation}

We now want to show how the direct subtraction and our unorthodox 
dimensional regularization are related. Taking Eq.~(\ref{eqn20}) for 
$n=3$ with a subtraction at the origin, one obtains
\begin{equation}\label{eqn25}
\Pi(p)=\int_0^\infty \left(\frac{\sin(px)}{px}-1\right) 
  \frac{e^{-Mx}}{(4\pi)^2x}(\mu^2 x^2)^\epsilon dx
\end{equation}
which is UV-finite even for $\epsilon=0$ because there is no singularity at 
the origin. For practical computations it is convenient to keep the factor 
$(\mu^2x^2)^\epsilon$ in the integrand 
since this factor gives a meaning to each of the two 
terms in the round brackets in Eq.~(\ref{eqn25}) separately. 
Then the direct 
computation gives
\begin{eqnarray}\label{eqn26}
\Pi(p)&=&\int_0^\infty\left(\frac{\sin(px)}{px}-1\right) 
  \frac{e^{-Mx}}{(4\pi)^2x}(\mu^2x^2)^\epsilon dx\nonumber\\
  &=&\frac{\Gamma(-1+2\epsilon)}{2ip(4\pi)^2}
  \left[(M-ip)^{1-2\epsilon}-(M+ip)^{1-2\epsilon}\right]\mu^{2\epsilon}
  -\frac{\Gamma(2\epsilon)}{(4\pi)^2}\left(\frac\mu{M}\right)^{2\epsilon}
  \qquad\nonumber\\
  &=&-\frac1{32\pi^2}\left\{\frac{M}{ip}\ln\left(\frac{M+ip}{M-ip}\right)
  +\ln\left(\frac{M^2+p^2}{M^2}\right)\right\}.
\end{eqnarray}
The poles cancel in this expression and the arbitrary scale $\mu$ changes to 
$M$. This corresponds to a transition from MS-type of renormalization schemes 
to a momentum subtraction scheme (with subtraction at the origin in
this particular case). Since the spectral density $\rho(s)$ is 
finite, it can be computed using any regularization scheme as can be seen 
from Eqs.~(\ref{eqn23}) and~(\ref{eqn26}). 

We mention that in the three-dimensional case the spectral density $\rho(s)$
can also be found for general values of $n$ by the traditional methods 
since the three-dimensional case is sufficiently simple. The use of the 
convolution equation needed for the evaluation keeps one in the same class 
of functions, i.e.\ polynomials in the variable $\sqrt{s}$ divided by 
$\sqrt{s}$~\cite{melone20}. The general form of the convolution equation in 
$D$-dimensional space-time reads
\begin{equation}\label{eqn27}
\Phi_n(s)=\int\Phi_k(s_1)\Phi_p(s_2)\Phi_2(s,s_1,s_2)ds_1ds_2,\qquad k+p=n.
\end{equation}
For the particular case of three-dimensional space-time the kernel 
$\Phi_2(p^2,m_1^2,m_2^2)$ is given by
\begin{equation}
(2\pi)^2\Phi_2(p^2,m_1^2,m_2^2)=\int\delta(k^2-m_1^2)
  \delta((p-k)^2-m_2^2)d^3k
\end{equation}
or, explicitly, 
\begin{equation}\label{eqn28}
\Phi_2(s,s_1,s_2)=\frac1{8\pi\sqrt s}\theta(s-(\sqrt{s_1}+\sqrt{s_2})^2).
\end{equation}
Eq.~(\ref{eqn28}) can be seen to be the two-particle phase space in 
three dimensions (cf.\ Eq.~(\ref{eqn22})). This is a rather simple example. 
However, our technique retains its efficiency for large $n$. It is also 
applicable for odd $D$ other than $3$, say $D=5$. Namely, the propagator in 
five-dimensional space-time reads ($\lambda_0=3/2$)
\begin{equation}
D(x,m)\rightarrow D_5(x,m)=\frac{(mx)^{3/2}K_{3/2}(mx)}{(2\pi)^{5/2}x^3}
  =\frac{e^{-mx}}{8\pi^2 x^3}(1+mx)
\end{equation} 
which assures that the integration in Eq.~(\ref{eqn07}) can be performed in 
terms of elementary functions (powers and logarithms) again.

We list some potential applications of the general results obtained in this 
subsection for odd-dimensional space-time. In three space-time dimensions 
our results can be used to compute phase space integrals for particles in 
jets where the momentum along the direction of the jet is 
fixed~\cite{melone21}. Another application can be found in three-dimensional 
QCD which emerges as the high temperature limit of the ordinary theory of 
strong interactions for the quark-gluon plasma (see 
e.g.~\cite{melone9,melone22,melone23,melone24}). Three-dimensional models 
are also used to study the question of dynamical mass generation and the 
infrared structure of the models of quantum field 
theory~\cite{melone25,melone26,melone27}. A further theoretical application 
consists in the investigation of properties of baryons in the limit of 
infinite number of colors $N_c\rightarrow\infty$ where one has to take into 
account the spin structure of internal lines. Note that particular models 
of different space-time dimensions are very useful because their properties 
may be simpler and may thus allow one to study general features of the 
underlying field theory. For example, in six-dimensional space-time the 
simplest model of quantum field theory $\phi^3$ is asymptotically free and 
can be used for simulations of some features of QCD. Though 
five-dimensional models are less popular than others, still there are useful 
applications for Yang-Mills theory in five-dimensional space-time where the 
UV structure of the models can be analyzed~\cite{melone28}.

We conclude this subsection by noting that the $x$-space techniques allow 
one to compute water melon diagrams in closed form in terms of elementary 
functions as long as one is dealing with odd-dimensional space-times. The 
resulting expressions are rather simple and can be directly used for 
applications. Having the complete formulas at hand, there is no need to 
expand in the parameters of the diagram such as masses or external 
momentum. In even number of space-time dimensions there are no closed form 
solutions in the general case. In this case it is useful to study some 
limiting configurations. In the next subsection we briefly formulate 
several limits for water melon diagrams in the Euclidean domain that can 
be computed analytically with the help of standard integrals given in 
textbooks~\cite{melone29}.   

\subsection{Limiting configurations and expansions\\in the Euclidean domain} 
What are the advantages of our method, especially for even dimensions? We 
shall find that, compared to existing approaches, our method results in 
great simplifications in computing the polarization function in the 
Euclidean domain. While the basic representation in Eq.~(\ref{eqn07}) can 
always be used for numerical evaluations, some analytical results can be 
obtained for particular choices of the parameters in the diagram (i.e.\ 
masses and external momenta). Different regimes can be considered and some 
cases can be explicitly done in closed form.

The simplest case is the limit of one large mass and all other masses being 
small. For the polarization function in the Euclidean domain this limit is 
easy to compute because of the simplicity of the $x$-space representation 
and the high speed of convergence of the ensuing numerical procedures. But 
this special limit can also be done analytically. When expanding the 
propagators in the limit of small masses one encounters powers of $x$ and 
$\ln(mx)$. The remaining functions are the weight function (the Bessel 
function) and the propagator of the heavy particle with the large mass 
$M_h$ which is given by the McDonald function. The general structure of the 
terms in the series expansion that contribute to $\Pi(p)$ is given by 
\begin{equation}\label{eqn29}
2\pi^{\lambda+1}\int_0^\infty\left(\frac{px}2\right)^{-\lambda}
  J_\lambda(px)K_\nu(M_hx)x^{2\rho}\ln^j(mx)dx,\quad\rho,j\ge 0.
\end{equation}
The integrations in~(\ref{eqn29}) can be done in closed form by using the 
basic integral representation
\begin{eqnarray}
\lefteqn{\int_0^\infty x^\mu J_\lambda(px)K_\nu(M_hx)dx}\nonumber\\
  &=&\frac{p^\lambda\Gamma((\lambda+\mu+\nu+1)/2)
  \Gamma((\lambda+\mu-\nu+1)/2)}{2^{1-\mu}M_h^{\lambda+\mu+1}
  \Gamma(\lambda+1)}\nonumber\\&&\qquad_2F_1\left((\lambda+\mu+\nu+1)/2,
  (\lambda+\mu-\nu+1)/2;\lambda+1;-p^2/M_h^2\right)
\end{eqnarray}
where $_2F_1(a,b;c;z)$ is the usual hypergeometric 
function~\cite{melone29}. The corresponding integrals with integer powers 
of logarithms ($j>0$ in Eq.~(\ref{eqn29})) can be obtained by 
differentiation with respect to $\mu$. Note that the maximal power of the 
logarithm is determined by the number of light propagators and does not 
increase with the order of the expansion: any light propagator contains only 
one power of the logarithm as follows from the expansion of the McDonald 
function $K_\nu(\xi)$ at small $\xi$~\cite{melone29}.

The basic formula Eq.~(\ref{eqn07}) is also well suited for finding
$p^2$-derivatives of the polarization function $\Pi(p)$. The values of the 
polarization function and its derivatives at $p^2=0$ can be easily obtained. 
The convenience of a $p^2$ expansion is demonstrated by making use of the 
basic formula for differentiating the Bessel function,
\begin{equation}\label{eqn30}
\frac{d^k}{d(p^2)^k}\left(\frac{px}2\right)^{-\lambda}J_\lambda(px)
  =(-x^2)^k\left(\frac{px}2\right)^{-\lambda-k}J_{\lambda+k}(px).
\end{equation}
Note that differentiation results in an expression which has the same 
functional structure as the original function. This is convenient for 
numerical computations. Note that sufficiently high order derivatives 
become UV-finite (operationally it is clear since the subtraction 
polynomial vanishes after sufficiently high derivatives are taken). This can 
also be seen explicitly from Eq.~(\ref{eqn30}) where high powers of $x^2$ 
suppress the singularity of the product of propagators at small $x$.

Mass corrections to the large $p^2$ behaviour in the Euclidean domain (an 
expansion in $m_i^2/p^2$) are obtained by expanding the massive propagators 
in terms of masses $m_i$ under the integration sign. The final integration 
is performed by using the formula
\begin{equation}
\int_0^\infty x^\mu J_\lambda(px)dx=2^\mu p^{-\mu-1}
  \frac{\Gamma\left({\lambda+\mu+1\over 2}\right)}
{\Gamma\left({\lambda-\mu+1\over 2}\right)}.
\end{equation}
Note that all these manipulations are straightforward and can be easily 
implemented in a system of symbolic computations. Some care is necessary, 
though, when poles of the $\Gamma$-function are encountered
which reflect the appearance of artificial infrared singularities. The 
corresponding framework for dealing with such problems is well known 
(see e.g.~\cite{melone30,melone31}). 

\section{Analytic continuation in momentum space}
Now we consider an explicit analytic continuation in the complex $p^2$ plane 
and locate the discontinuity of the polarization function which is nothing 
but the spectral density $\rho(s)$. Note that the spectral density 
represents $n$-particle phase space for the water melon configuration.
 
Again we use the basic formula given by Eq.~(\ref{eqn07}) for analytic 
continuation. The spectral density reads
\begin{eqnarray}\label{eqn31}
\rho(s)&=&\frac{i}{2\pi}\int_0^\infty
  \left(\frac{2\pi\xi}s\right)^{\lambda+1}J_\lambda(\xi)\Bigg[
  e^{-i\pi(\lambda+1)}\prod_{i=1}^n\frac14\left(\frac{m_i\sqrt s}{2\pi\xi}
  \right)^\lambda e^{i\pi(\lambda+1/2)}H^{(1)}_\lambda
  \left(\frac{m_i\xi}{\sqrt s}\right)\nonumber\\&&\qquad\qquad
  -e^{i\pi(\lambda+1)}\prod_{i=1}^n\frac14\left(\frac{m_i\sqrt s}{2\pi\xi}
  \right)^\lambda e^{-i\pi(\lambda+1/2)}H^{(2)}_\lambda
  \left(\frac{m_i\xi}{\sqrt s}\right)\Bigg]d\xi.
\end{eqnarray} 
The analytic continuation is performed using
the result
\begin{equation}
K_\lambda(z)=\frac{\pi i}2e^{\frac{\pi}2\lambda i}H^{(1)}_\lambda(iz)   
\end{equation}
where $H^{(1,2)}_\lambda(z)$ are the Hankel functions, 
$H^{(1)}_\lambda(z)=(H^{(2)}_\lambda(z))^*$ for real 
$z,\lambda$~\cite{melone29}. We have thus obtained an explicit 
representation for the spectral density in terms of a one-dimensional 
integral representation. Using Eq.~(\ref{eqn31}) one can easily discuss 
special mass configurations. It is understood that the UV subtraction has 
been performed in Eq.~(\ref{eqn31}), i.e.\ the kernel 
$(\xi/2)^{-\lambda}J_\lambda(\xi)$ is substituted with the subtracted one 
$[.]_N$ from Eq.~(\ref{eqn09}).

A remark about the numerical evaluation of the integrals in Eq.~(\ref{eqn31})
is in order. As our main purpose is to create a practical tool for 
evaluation of the water melon class of diagrams, we do not insist on an 
analytical evaluation of the integrals in Eq.~(\ref{eqn31}). This issue
will be discussed in more detail in Sec.~4. The one-dimensional integral 
representation in Eq.~(\ref{eqn31}) is simple enough for further processing. 
The evaluation of the integral in Eq.~(\ref{eqn31}) is, however, not always 
straightforward. The integrand contains highly oscillating functions that 
require some care in the numerical treatment. This is to be expected since 
the discontinuity, or the spectral density, is a distribution rather than a 
smooth function. However, because the analytic structure and the asymptotic 
behaviour of the integrand in Eq.~(\ref{eqn31}) is completely known, the 
numerical computation of $\rho(s)$ can be made reliable and fast in domains 
where $\rho(s)$ is smooth enough, in particular far from threshold. One 
recipe is to extract the oscillating asymptotics first and then to perform 
the integration analytically, or to integrate the oscillating asymptotics 
numerically using integration routines that have special options for the 
treatment of oscillatory integrands. Both ways were checked in simple 
examples with reliable results. The remaining non-oscillating part is a 
slowly changing function which can be integrated numerically without 
difficulties. With this extra care the integration can be easily made safe, 
reliable and fast even for an average personal computer. We mention that we 
have checked our general numerical procedures in three-dimensional 
space-time ($D_0=3$) where exact results are available (see Sec.~2).

As an example of the efficiency of our technique we take Eq.~(\ref{eqn31}) 
to recompute the spectral density of the $n=3$ water melon diagram for
three-dimensional space-time. The Hankel function for indices $j+1/2$ (or 
for $D_0=2j+1$) is a finite combination of powers and an exponential which 
makes possible the explicit computation of the integral in Eq.~(\ref{eqn31}). 
In fact, for this case one has
\begin{equation}\label{eqn32}
\rho(s)=-\frac1\pi\int_0^\infty\left(\frac{\sin\xi}\xi-1\right)
  \sin\left(\frac{M\xi}{\sqrt{s}}\right)\frac{d\xi}{(4\pi)^2\xi}
  =\frac{\sqrt s-M}{32\pi^2\sqrt s}\,\theta(s-M^2).
\end{equation}
This form coincides with the explicit formulas given by Eqs.~(\ref{eqn21}) 
and~(\ref{eqn22}). The generalization to higher $n$ includes only algebraic 
manipulations. The necessary integrations corresponding to the one in 
Eq.~(\ref{eqn32}) are performed by moving the contour into the complex 
$\xi$-plane and regularizing the singularity at the origin by an infinitely 
small shift $\pm i0$. Then closing the contour in the upper or lower 
semi-plane according to the sign of regularization one finds the integral 
by computing the residue at the origin. Note that an explicit subtraction 
is kept in Eq.~(\ref{eqn32}).

The results for the spectral density for $n=3$ and~$4$ can be obtained 
directly by making use of traditional techniques as well. One obtains a 
one-dimensional integral representation for $n=3$ and a two-dimensional 
integral representation for $n=4$~\cite{melone20}. For larger $n$, however, 
the corresponding technique of convolution includes many-fold integrals and 
the corresponding recursion relation (\ref{eqn27}) is not convenient for 
applications. This has to be contrasted with the one-dimensional integral 
representation in Eq.~(\ref{eqn31}) derived here which allows one to compute 
the spectral density for the class of water melon diagrams with large 
number of internal lines in any number of space-time dimensions.

\section{Integral transformation in configuration space} 
The analytic structure of the correlator $\Pi(x)$ (or the spectral density 
of the corresponding polarization operator) can be determined directly in 
configuration space without having to compute its Fourier transform first.
The dispersion representation (or the spectral decomposition) of the 
polarization function in configuration space has the form 
\begin{equation}\label{eqn33}
\Pi(x)=\int_0^\infty\rho(m^2)D(x,m)dm^2
\end{equation}
where for this section we switch to a notation where $\sqrt s=m$. This 
representation was used for sum rules applications 
in~\cite{melone32,melone33} where the spectral density for the two-loop 
sunset diagram was found in two-dimensional space-time~\cite{melone34}. 
With the explicit form of the propagator in configuration space given by
Eq.~(\ref{eqn02}), the representation in Eq.~(\ref{eqn33}) turns into a 
particular example of the Hankel transform, namely the 
$K$-transform~\cite{melone35,melone36}. Up to inessential factors of $x$ 
and $m$, Eq.~(\ref{eqn33}) reduces to the generic form of the 
$K$-transform for a conjugate pair of functions $f$ and $g$,
\begin{equation}
g(y)=\int_0^\infty f(x)K_\nu(xy)\sqrt{xy}\,dx.
\end{equation}
The inverse of this transform is known to be given by 
\begin{equation}\label{eqn34}
f(x)=\frac1{\pi i}\int_{c-i\infty}^{c+i\infty}g(y)I_\nu(xy)\sqrt{xy}\,dy
\end{equation}
where $I_\nu(x)$ is a modified Bessel function of the first kind and the 
integration runs along a vertical contour in the complex plane to the right 
of the right-most singularity of the function $g(y)$~\cite{melone36}. In 
order to obtain a representation for the spectral density $\rho(m^2)$ of a 
water melon diagram in general $D$-dimensional space-time one needs to 
apply the inverse $K$-transform to the particular case given by 
Eq.~(\ref{eqn33}). One has 
\begin{equation}\label{eqn35}
m^\lambda\rho(m^2)=\frac{(2\pi)^\lambda}{i}
  \int_{c-i\infty}^{c+i\infty}\Pi(x)x^{\lambda+1}I_\lambda(mx)dx.
\end{equation}
The inverse transform given by Eq.~(\ref{eqn35}) solves the problem of 
determining the spectral density of water melon diagrams by reducing it to 
the computation of a one-dimensional integral for the general class of 
water melon diagrams with any number of internal lines and different masses. 
Compared to the general solution given by Eq.~(\ref{eqn31}) the above form 
is simpler. Below we discuss some technicalities concerning the efficient 
evaluation of the contour integral in the representation given by 
Eq.~(\ref{eqn35}). The analytic structure of the correlator in 
Eq.~(\ref{eqn01}) is now explicit using the representation given by 
Eq.~(\ref{eqn35}) and exhibits the distribution nature of the spectral 
density $\rho(m^2)$ as shown below.

Note that the expression given by Eq.~(\ref{eqn01}) can have non-integrable
singularities at small $x$ for a sufficiently large number of propagators 
when $D>2$~\cite{melone16}. Therefore the computation of its Fourier 
transform requires regularization or subtractions~\cite{melone37}. The 
spectral density itself is finite (the structure of the water melon 
diagrams is very simple and there are no subdivergences when one employs 
a properly defined $R$-operation~\cite{melone16}) and thus requires no 
regularization. In the more traditional approach of direct analytic 
continuation in momentum space the explicit representation for the spectral 
density is given by Eq.~(\ref{eqn31}) where one has taken the 
discontinuity of the Fourier transform across the physical cut for 
$p^2=-m^2\pm i0$. This is an alternative representation of the spectral 
density, and in some instances the latter representation can be more 
convenient for numerical treatment.

In the following we present some explicit examples of applying the 
technique of computing the spectral density of water melon diagrams on the
basis of integral transforms in configuration space.

\subsection{One-loop case}
First a remark about the mass degenerate one-loop case is in order. All 
necessary integrals (both for the direct and the inverse $K$-transform) 
involve no more than the product of three Bessel functions which can be 
found in a standard collection of formulas for special functions (see 
e.g.~\cite{melone29}). The spectral density in $D$-dimensional space-time 
(for two internal lines with equal masses $m_0$) can be computed to be
\begin{equation}\label{eqn36}
\rho(m^2)=\frac{(m^2-4m_0^2)^{\lambda-1/2}}{2^{4\lambda+1}\pi^{\lambda+1/2}
  \Gamma(\lambda+1/2)m},\qquad m>2m_0.
\end{equation}
This formula is useful since it can be used to test the limiting cases of 
more general results.

The corresponding spectral density for the nondegenerate case with two 
different masses $m_1$ and $m_2$ reads
\begin{equation}\label{eqn37}
(2\pi)^{2\lambda+1}\rho(m^2)=\frac{\Omega_{2\lambda+1}}{4m}
  \left(\frac{(m^2-m_1^2-m_2^2)^2-4m_1^2m_2^2}{4m^2}\right)^{\lambda-1/2}
  \hspace{-24pt},\qquad m>m_1+m_2,
\end{equation}
where 
\begin{equation}
\Omega_d=\frac{2\pi^{d/2}}{\Gamma(d/2)}
\end{equation}
is a volume of a unit sphere in $d$-dimensional space-time.
Note the identity
\begin{equation}
(m^2-m_1^2-m_2^2)^2-4m_1^2m_2^2
  =\left[m^2-(m_1+m_2)^2\right]\left[m^2-(m_1-m_2)^2\right]
\end{equation}
which immediately allows one to locate the two-particle threshold.

\subsection{Odd-dimensional case}
For odd-dimensional space-time the representation in Eq.~(\ref{eqn33}) 
reduces to the ordinary Laplace transformation. To obtain the spectral 
density (the function $f(x)$ in this particular example) one can use 
Eq.~(\ref{eqn34}). For energies below threshold it is possible to close the 
contour of integration to the right. With the appropriate choice of the 
constant $c$ as specified above, the closed contour integration gives zero 
due to the absence of singularities in the relevant domain of the right 
semi-plane. By closing the contour of integration to the left and keeping 
only that part of the function $I_\nu(z)$ which is exponentially falling for 
${\rm Re}(z)<0$ one can obtain another convenient integral representation 
for the spectral density when the energy is above threshold. The only 
singularities within the closed contour are then poles at the origin (in 
odd-dimensional space-time) and the evaluation of the integral can be done 
by determining the corresponding residues. These are purely algebraic 
manipulations, the simplicity of which also explain the simplicity of the 
computations in odd-dimensional space-time. For a small number of internal 
lines $n$ the spectral density can also be found by using the convolution 
formulas for the spectral densities of a smaller number of particles (see 
e.g.~\cite{melone20}). For large $n$ the computations described 
in~\cite{melone20} become quite cumbersome and the technique suggested in 
the present paper is much more convenient. 

As an example for the odd-dimensional case we present calculations in three 
dimensions. The dispersion representation for three-dimensional space-time 
has the following form
\begin{equation}\label{eqn38}
\Pi(x)=\int_0^\infty\rho(m^2)D_3(x,m)dm^2
  =\int_0^\infty\rho(m^2){e^{-mx}\over 4\pi x}dm^2
\end{equation}
with the three-dimensional scalar propagator
\begin{equation}\label{eqn39}
D_3(x,m)=\frac{\sqrt{mx}K_{1/2}(mx)}{(2\pi)^{3/2}x}=\frac{e^{-mx}}{4\pi x}.
\end{equation}
One can invert Eq.~(\ref{eqn38}) and obtains
\begin{equation}\label{eqn40}
2m\rho(m^2)=\frac1{2\pi i}\int_{c-i\infty}^{c+i\infty}4\pi x\Pi(x)e^{mx}dx
\end{equation}
which is a special case of Eqs.~(\ref{eqn34}) and~(\ref{eqn35}) with 
\begin{equation}
I_{\frac12}(z)=\sqrt{\frac2{\pi z}}\sinh(z)
\end{equation}
where one only needs to retain the $e^z$ piece in the hyperbolic sine 
function. The solution given by Eq.~(\ref{eqn40}) has the appropriate 
support as a distribution or, equivalently, as an inverse Laplace transform. 
It vanishes for $m<M=\sum_{i=1}^nm_i$\quad since the contour of integration 
can be closed to the right where there are no singularities of the 
integrand. Recall that for large $x$ with ${\rm Re}(x)>0$ the asymptotic 
behaviour of the polarization function $\Pi(x)$ is governed by the sum of 
the masses of the propagators and reads
\begin{equation}
\Pi(x)\sim\exp{(-xM)}.
\end{equation}
For $m>M$ one can close the contour to the left in the complex $x$-plane 
and then the only singularities of $\Pi(x)$ are the poles at the origin of 
$\Pi(x)$ since $\Pi(x)$ is a product of the propagators of the form of 
Eqs.~(\ref{eqn01}) and~(\ref{eqn39}). The integration in Eq.~(\ref{eqn40}) 
then reduces to finding the residues of the poles at the origin. Indeed,
\begin{equation}
\Pi(x)=\prod_{i=1}^nD_3(x,m_i)={e^{-Mx}\over(4\pi x)^n},
\end{equation}
and Eq.~(\ref{eqn40}) gives an explicit representation of the spectral
density through the polarization function in $x$-space,
\begin{equation}\label{eqn41}
2m\rho(m^2)=\frac1{2\pi i}\int_{c-i\infty}^{c+i\infty}4\pi x\Pi(x)e^{mx}dx
  =\frac1{2\pi i(4\pi)^{n-1}}\int_{c-i\infty}^{c+i\infty}
  \frac{e^{(m-M)x}}{x^{n-1}}dx.
\end{equation}
In closing the contour of integration to the left one computes the residue 
at the origin and obtains ($n>1$)
\begin{equation}\label{eqn42}
2m\rho(m^2)=\frac{(m-M)^{n-2}}{(4\pi)^{n-1}(n-2)!}\theta(m-M). 
\end{equation}
Eq.~(\ref{eqn42}) coincides with the expression (\ref{eqn24}). This also 
explains the simplicity of the structure of the spectral density in odd 
numbers of dimensions of space-time when traditional means are 
used~\cite{melone37}. In five-dimensional space-time Eq.~(\ref{eqn40})
is applicable almost without any change because the propagator now reads
\begin{equation}
D_5(x,m)=\frac{(mx)^{3/2}K_{3/2}(mx)}{(2\pi)^{5/2}x^3}
  =\frac{e^{-mx}}{8\pi^2x^3}(1+mx).
\end{equation}
Compared to the three-dimensional case the only additional complication 
is that the order of the pole at the origin is changed and
that one now has a linear combination of terms instead of the simple 
monomial in three dimensions.

\subsection{Even-dimensional case}
For even-dimensional space-time the analytic structure of $\Pi(x)$ in 
Eq.~(\ref{eqn01}) is more complicated. There is a cut along the negative 
axis in the complex $x$-plane which prevents a straightforward evaluation 
by simply closing the contour of integration to the left (with 
${\rm Re}(x)<0$). The discontinuity along the cut is, however, well known 
and includes only Bessel functions that appear in the product of 
propagators for the polarization function. Therefore the 
representation~(\ref{eqn35}) is essentially equivalent to the direct
analytic continuation of the Fourier transform~\cite{melone37} but may be 
more convenient for numerical treatment because there is no oscillating 
integrand in~(\ref{eqn35}).
 
In even number of dimensions one is thus dealing with a genuine 
$K$-transform. We discuss in some detail the important case of 
four-dimensional space-time. For $D=4$ ($\lambda=1$), Eqs.~(\ref{eqn02}) 
and~(\ref{eqn33}) give
\begin{equation}
\Pi(x)=\int\rho(m^2)D_4(x,m)dm^2=\int\rho(m^2)
  \frac{mxK_1(mx)}{4\pi^2x^2}dm^2,
\end{equation}
and Eq.~(\ref{eqn35}) is written as
\begin{equation}\label{eqn43}
2m\rho(m^2)=\frac1{\pi i}\int_{c-i\infty}^{c+i\infty}
  4\pi^2x^2\Pi(x)I_1(mx)dx.
\end{equation}
All remarks about the behaviour at large $x$ apply here as well. However, 
the structure of singularities is more complicated than in the 
odd-dimensional case. In addition to the poles at the origin there is a cut 
along the negative axis that renders the computation of the spectral 
density more difficult. The cut arises from the presence of the functions 
$K_1(m_ix)$ in the polarization function $\Pi(x)$. Also the asymptotic 
behaviour of the function $I_1(z)$ is more complicated than that of 
$I_{1/2}(z)$. In particular the extraction of the exponentially falling 
component on the negative real axis is not straightforward. Incidentally, the
fall-off behaviour of the function $I_1(z)$ on the negative real axis can
be taken as an example of Stokes' phenomenon of asymptotic expansions 
(see e.g.~\cite{melone12}). While the analytic structure of the 
representation is quite transparent and the integration can be performed 
along a contour in the complex plane, there are some subtleties when one 
wants to obtain a convenient form for numerical treatment analogous 
to odd-dimensional case ~\cite{melone37}.

After closing the contour to the left (for $m>M$) using the appropriate 
part of the function $I_1(z)$ we obtain
\begin{eqnarray}\label{eqn44}
\lefteqn{i\pi\int_{c-i\infty}^{c+i\infty}x^2\Pi(x)I_1(mx)dx}\\ 
  &=&-\int_\epsilon^\infty
  r^2\left(\Pi(e^{i\pi}r)+\Pi(e^{-i\pi}r)\right)K_1(mr)dr
  +2\int_\epsilon^\infty r^2\Pi(r)K_1(mr)dr\nonumber\\&&
  +\int_{C_-}z^2\Pi(z)(i\pi I_1(mz)+K_1(mz))dz
  +\int_{C_+}z^2\Pi(z)(i\pi I_1(mz)-K_1(mz))dz\nonumber
\end{eqnarray}
for the quantity entering Eq.~(\ref{eqn43}). The contours $C_+$ and $C_-$ 
are semi-circles of radius $\epsilon$ around the origin in the upper and 
lower complex semi-plane, respectively (see Fig.~3).
\begin{figure}\begin{center}
\epsfig{figure=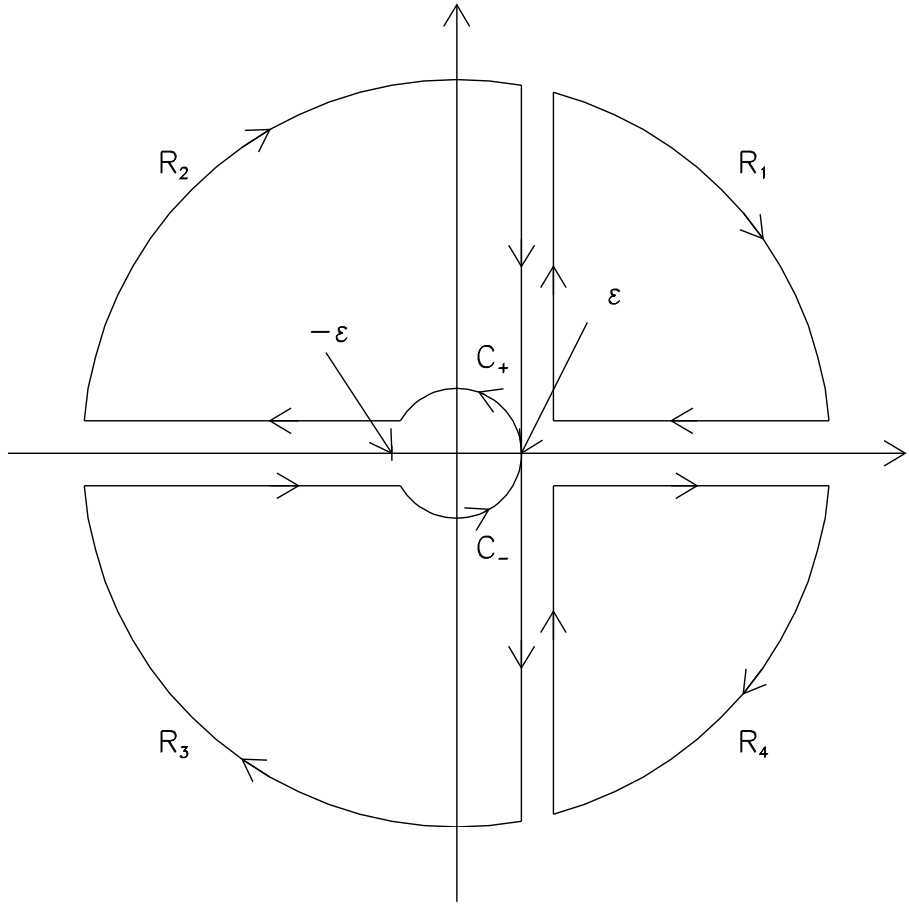, height=12truecm, width=12truecm}
\caption{\label{fig3}Integration contours used in the evaluation of 
Eq.~(\ref{eqn44}). $R_1$, $R_2$, $R_3$ and $R_4$ are segments of a circle 
arround the origin where the radius of the circle is taken to infinity}
\end{center}\end{figure}
For practical evaluations of $\Pi(e^{\pm i\pi}r)$ the following rule for the
analytic continuation of the McDonald functions is used,
\begin{equation}
K_1(e^{\pm i\pi}\xi)=-K_1(\xi)\mp i\pi I_1(\xi),\qquad\xi=mr>0.   
\end{equation}

Some comments are in order. The polarization function $\Pi(z)$ at 
$z=e^{\pm i\pi}r$ is proportional to the product of propagators of the form 
\begin{equation}
D_4(e^{\pm i\pi}r,m_i)
  \sim\frac{m_i}r\left[K_1(m_ir)\pm i\pi I_1(m_ir)\right].
\end{equation}
It is clear from this equation that the leading singular contribution
proportional to the product of $K_1(m_ir)$ cancels in the sum in 
Eq.~(\ref{eqn44}). Also the next-to-leading singular term disappears 
because of different signs in the product. Recall that the small $\xi$ 
behaviour of the functions $K_1(\xi)$ and $I_1(\xi)$ is given by
\begin{equation}
K_1(\xi)=\frac1\xi+O(\xi\ln\xi),\quad I_1(\xi)=\frac\xi2+O(\xi^3).
\end{equation}

Let us add a few remarks on the final representation Eq.~(\ref{eqn44}) 
which is in a suitable form for numerical integration. We have introduced 
an auxiliary regularization in terms of a circle of finite radius 
$\epsilon$ which runs around the origin with its pole-type singularities. 
The spectral density is independent of $\epsilon$, and the parameter 
$\epsilon$ completely cancels in the full expression for the spectral 
density as given by Eq.~(\ref{eqn44}). This must be so since the spectral 
density is finite for the class of water melon diagrams. Eq.~(\ref{eqn44}) 
contains no oscillatory integrands (cf.\ Eq.~(\ref{eqn31})), and the 
integration can safely be done numerically. Thus Eq.~(\ref{eqn44}) is a 
useful alternative representation for the spectral density. In practice 
the integration over the semi-circles is done by expanding the integrand 
in $z$ for small $z$ and keeping only terms singular in $\epsilon$. The 
expansion requires only a finite number of terms and is a purely algebraic 
operation. Then the singularity in $\epsilon$ exactly cancels against those 
of the remaining integrals. This cancellation can also be done analytically 
leaving well defined and smooth integrands for further numerical treatment.

Even if the full computation of the spectral function in the 
even-dimensional case described here is straightforward it is nevertheless 
cumbersome. In order to exhibit the essential points in this calculation we 
illustrate it with a simple and instructive example. Consider the 
calculation of the following integral
\begin{equation}
\int_{c-i\infty}^{c+i\infty}\frac{\ln z}{z^2}e^zdz
\end{equation}
which is rather close in structure to the real case. Due to the singularity 
at the origin one has to treat the vicinity of the origin carefully. We 
proceed by closing the contour to the left
\begin{eqnarray}\label{eqn45}
\int_{c-i\infty}^{c+i\infty}\frac{\ln z}{z^2}e^zdz
  &=&\int_{C_\epsilon}\frac{\ln z}{z^2}e^zdz
  -\int_\epsilon^\infty\frac{2i\pi}{x^2}e^{-x}dx \nonumber \\ 
  &=&-2\pi i\int_\epsilon^{\infty}{e^{-x}dx\over x^2}
  +i\int_{-\pi}^{\pi}{\ln\epsilon+i\phi\over \epsilon}(1+\epsilon
  e^{i\phi})e^{-i\phi}d\phi \nonumber \\
  &=&-2\pi i\int_\epsilon^{\infty}{e^{-x}dx\over x^2}
  +i\int_{-\pi}^{\pi}
  \left({\ln\epsilon\over \epsilon} e^{-i\phi}+{i\phi\over \epsilon}
  e^{-i\phi}+\ln\epsilon+i\phi\right)d\phi\nonumber \\
  &=&-2\pi i\int_\epsilon^{\infty}{e^{-x}dx\over x^2}
  +{\ln\epsilon\over \epsilon}0+{2\pi i\over \epsilon}
  +2\pi i\ln\epsilon+0\nonumber\\
  &=&-2\pi i\left(\int_\epsilon^{\infty}{e^{-x}dx\over x^2}
  -{\frac{1}{\epsilon}}-\ln\epsilon\right)\,.
\end{eqnarray}
The combination in the brackets of the last equation remains finite as 
$\epsilon\rightarrow 0$. Let us now consider this limit in more detail. 
First we split the integration into two parts from $\epsilon$ to $1$ and 
from $1$ to infinity,
\begin{equation}\label{eqn46}
\int_\epsilon^{\infty}{e^{-x}dx\over x^2}
  =\int_1^{\infty}{e^{-x}dx\over x^2}
  +\int_\epsilon^{1}{e^{-x}dx\over x^2}.  
\end{equation}
Then the first integral is just a number which can be found numerically 
with high precision. In the second integral we can expand the exponent in 
the integrand and find
\begin{equation}
\int_\epsilon^{1}{e^{-x}dx\over x^2} =
\int_\epsilon^{1}{dx\over x^2}(1-x+\frac{x^2}{2}-\ldots)=
\frac{1}{\epsilon}-1+\ln{\epsilon}+\frac{1}{2}(1-\epsilon)+\ldots.
\end{equation}
The singularity has the correct form and the finite series converges well.
If one takes a value $0.1$ instead of $1$ for the splitting point in 
Eq.~(\ref{eqn46}) the convergence of the finite series will be very fast. 
This procedure would be used for practical integration in a realistic case.

Now we give an exact answer for our simple example. After two integrations 
by parts we have 
\begin{equation}
\int_\epsilon^{\infty}{e^{-x}dx\over x^2}
  =e^{-\epsilon}\left(\frac{1}{\epsilon}+\ln{\epsilon}\right)
  -\int_\epsilon^{\infty}e^{-x}\ln x dx.
\end{equation}
The last integral is finite at $\epsilon=0$ and for our purpose it suffices 
to compute it in this limit. The result is 
\begin{equation}
\int_0^{\infty}e^{-x}\ln x\,dx=-\gamma_E
\end{equation}
where $\gamma_E$ is again the Euler's constant. All these manipulations can 
be easily done with a symbolic program. For the original integral one finds
\begin{eqnarray}
-2\pi i\left(\int_\epsilon^{\infty}{e^{-x}dx\over x^2}
  -{\frac{1}{\epsilon}}-\ln\epsilon\right)&=&-2\pi i
  \left(e^{-\epsilon}\left(\frac{1}{\epsilon}+\ln{\epsilon}\right)
  +\gamma_E-{\frac1{\epsilon}}-\ln\epsilon\right)\nonumber\\
&=&2\pi i(1-\gamma_E)\quad\mbox{at}\quad\epsilon=0.
\end{eqnarray}
Thus finally
\begin{equation}
\int_{c-i\infty}^{c+i\infty}\frac{\ln z}{z^2}e^zdz=2\pi i(1-\gamma_E)\,.
\end{equation}
This concludes our discussion of how to treat subtractions in this 
simplified case. The generalization to Bessel functions is straightforward
(just expand near the origin and get expressions as in Eq.~(\ref{eqn45})).
Then the form of the subtraction term depends on the number $n$ of
propagators in a water melon diagram. Writing down an explicit expression 
for some $n$ is routine and we leave it to the interested user. All the 
required expansions can be performed by a symbolic manipulation program. 

Again for $n=3$ the results for the spectral density can be obtained 
directly by traditional means through the convolution equation~(\ref{eqn27}).
It leads to a one-dimensional integral representation for $n=3$. In this 
respect the representation in Eq.~(\ref{eqn44}) is of the same level of 
complexity while the convolution equation is even simpler because it 
includes only an integration over a finite interval. However, for $n=4$ 
the convolution equation leads to a two-dimensional integral 
representation~\cite{melone20}. For larger $n$ the corresponding technique of 
convolution includes many-fold integrals and the corresponding recursion 
relation in Eq.~(\ref{eqn27}) is not very convenient for applications. In 
the configuration space approach all formulas remain the same regardless of 
the number $n$ of internal lines and thus this approach must be considered 
to be superior to the more traditional momentum space approach.

For two-dimensional space-time the representation analogous to 
Eq.~(\ref{eqn44}) is simpler because there is no power singularity at the 
origin but only a logarithmic singularity which allows one to shrink the 
contour to a point (take the limit $\epsilon\rightarrow 0$). In this case 
we obtain
\begin{equation}
\rho(m^2)=\frac1\pi\int_0^\infty r(2\Pi(r)-\Pi(e^{i\pi}r)
  -\Pi(e^{-i\pi}r))K_0(mr)dr.
\end{equation}
For $D_0=2$ we present our results for the cases $n=2$ and $n=3$. For the 
one-loop case $n=2$ one has
\begin{equation}
\rho(m^2)=\frac1{2\pi}\int_0^\infty rI_0(m_1r)I_0(m_2r)K_0(mr)dr
\end{equation}
which can be integrated explicitly and results in~\cite{melone19}
\begin{equation}
\rho(m^2)=\frac1{2\pi m^2}\sum_{k,l=0}^\infty
  \left(\frac{(k+l)!}{k!l!}\right)^2\left(\frac{m_1^2}{m^2}\right)^k
  \left(\frac{m_2^2}{m^2}\right)^l.
\end{equation}
This series can of course also be directly obtained by expanding 
Eq.~(\ref{eqn37}).

For the case $n=3$ one obtains
\begin{eqnarray}\label{eqn47}
\rho(m^2)&=&\frac1{(2\pi)^2}\int_0^\infty r(K_0(m_1r)I_0(m_2r)I_0(m_3r)
  +I_0(m_1r)K_0(m_2r)I_0(m_3r)\nonumber\\&&\qquad\qquad
  +I_0(m_1r)I_0(m_2r)K_0(m_3r))K_0(mr)dr.
\end{eqnarray}
Again, this result can be obtained by an alternative method, i.e.\ by 
making use of the convolution equation given in Eq.~(\ref{eqn27}). The 
$\delta$-function $\Phi_1(m^2)=\delta(m^2-m_3^2)$ is the spectral density 
of the additional internal line with mass $m_3$. $\Phi_2(m^2)$ and 
$\Phi_2(m^2,m_1^2,m_2^2)$ are the one-loop spectral density given 
by Eq.~(\ref{eqn37}). In performing the convolution one obtains
\begin{equation}\label{eqn48}
\rho(m^2)=\frac1{(2\pi)^2}\int_{(m_1+m_2)^2}^{(m-m_3)^2}\frac{ds_1}
 {\sqrt{((s_1-m_1^2-m_2^2)^2-4m_1^2m_2^2)((m^2-s_1-m_3^2)^2-4s_1m_3^2)}}.
\end{equation}
The integrand in Eq.~(\ref{eqn48}) is singular at the end points and 
a numerical evaluation 
requires some care. Contrary to this, there are 
no problems when Eq.~(\ref{eqn47}) is used. 
With modern 
computer facilities, the integral representation given in 
Eq.~(\ref{eqn47}) is more suitable for a numerical evaluation than 
the form given by Eq.~(\ref{eqn48}).

The most direct and efficient way of numerical evaluation of the spectral 
density is the immediate use of Eq.~(\ref{eqn35}). After choosing the 
contour to be a straight vertical line in the complex $x$-plane with some 
positive $c$ and making change of the variable $x=c+iy$ the integration 
over $y$ is straightforward. For large number of lines $n$ in a water melon 
diagram the convergence of the integral becomes fast because the integrand 
falls off as $|y|^{-3(n-1)/2}$ at large $|y|$. One finds several correct 
decimal figures of the result rather easily using an ordinary personal 
computer. A good check for the quality of numerical results obtained in 
this way is their independence of $c$. The results must be the same for any 
particular choice of $c$. 

\subsection{Threshold behaviour}
Finally we return to the threshold behaviour of the spectral density. Using 
the results of the above analysis one arrives at expressions of the general 
form 
\begin{equation}\label{eqn49}
2m\rho(m^2)\sim\int^\infty K_1(mr)\prod_iZ_1(m_ir)dr
\end{equation}
where $Z_1(m_ir)$ is either $I_1(m_ir)$ or $K_1(m_ir)$. The convergence at 
large $r>0$ (at the upper limit of the integral) is controlled by the factor 
$\exp{(-(m-M)r)}$ as in Eq.~(\ref{eqn41}), and the corresponding expansions 
in the variable $m-M$ in the region $m\sim M$ can be constructed. As 
shown before, the spectral density for odd-dimensional space-time is known
exactly. Its threshold behaviour is easily extracted and one obtains
\begin{equation}
\rho(m^2)\approx\frac{(m-M)^{n-2}}{2m(4\pi)^{n-1}(n-2)!}
\end{equation}
at $m$ close to $M$.

In the realistic case of four-dimensional space-time the threshold behaviour 
of the spectral density can be inferred from 
Eq.~(\ref{eqn35}) (and, also, Eq.~(\ref{eqn49})). Substituting the 
asymptotic limit of modified Bessel functions at large arguments we infer 
from the simple dimensional considerations that 
\begin{equation}
\rho(m^2)\sim(m-M)^{(3n-5)/2}
\end{equation}
at $m$ close to $M$, where $n$ is the number of internal lines of the water 
melon diagram. One obtains $\rho(m^2)\sim\sqrt{m-M}$ for $n=2$ and 
$\rho(m^2)\sim(m-M)^2$ for $n=3$ at $m$ close to $M$. The threshold 
behaviour agrees with the one extracted from the explicit formula given by 
Eq.~(\ref{eqn36}) or with the threshold behaviour derived from the 
convolution equation in Eq.~(\ref{eqn27}). Other space-time dimensions can 
be analyzed along the same lines.

For the sake of completeness we briefly comment on the opposite limit 
for the spectral density when $m^2\rightarrow\infty$. In this limit the 
external energy is much larger than all masses. The limiting behaviour of 
this kinematic configuration can be found by utilizing the massless 
approximation for the correlator where analytical expressions are 
available. Subleading corrections can be obtained using known 
techniques~\cite{melone4}.

\section{Conclusion}
We have described a novel technique that reduces the computation of any 
water melon diagram to a simple one-dimensional integral with known 
functions in the integrand. Additional tensor and form factor structures 
can be easily included without modification of the basic formulas. 
Different regimes of behaviour with respect to mass/momentum expansions can 
be easily analyzed and the results can be found analytically if no more 
than two dimensionful parameters (one mass and momentum or two masses)
are treated exactly and other parameters are considered to be small and 
are treated in series expansions. Explicit analytical formulas (with 
even that last integration being performed in closed form) are given
for the case of any odd number of space-time dimensions. The analytic 
continuation to the Minkowskian region, i.e.\ to the positive $s$-axis has 
been performed and the discontinuity across the physical cut is explicitly 
given. This allows one to compute general $n$-particle phase space for any 
kind of particles both with different masses and Lorentz structures. The 
threshold behaviour of the spectral density can be easily investigated 
based on such a representation. In the even-dimensional case the final 
integration for an arbitrary water melon diagram was not done explicitly 
since we encountered products of Bessel functions which we were unable to 
integrate in a closed form. This forces one to use numerical integrations 
in the even-dimensional case if numerical results are desired for large 
$n$. However, the analytic structure of the solution for the diagram is 
fully determined which makes the numerical treatment rather reliable and 
fast. In this sense the computation of water melon diagrams has been 
converted to the routine procedure of getting numbers. Except for some 
special cases the final one-dimensional integral representation was not 
further reduced to known simpler functions. This is quite familiar from 
the theory of special functions where many special functions are 
represented through one-dimensional integrals which cannot be further 
reduced to known functions. In this sense our formulas completely solve 
the problem of computing the class of water melon diagrams.

Well, the water melon is finished and the sunset is over long ago.
So the next story has to wait until the next sunrise takes place.

\vspace{3mm}\noindent
{\large\bf Acknowledgments}\\[2mm]
We thank A.~Davydychev for useful comments on the literature related to the
subject of the present paper. We futher thank D.~Broadhurst for an 
interesting discussion. A.~Grozin has provided us with the program 
RECURSOR~\cite{melone18} which allowed us to obtain the analytical result in 
Eq.~(\ref{eqn18}). The work is supported in part by the Volkswagen 
Foundation under contract No.~I/73611. A.A.~Pivovarov is supported in part 
by the Russian Fund for Basic Research under contracts Nos.~96-01-01860 and 
97-02-17065.

\end{document}